\documentclass{amsart}

\usepackage{color}  
\usepackage{centernot} 
\usepackage{graphicx} 
\usepackage{epstopdf} 
\usepackage{amsaddr} 

\usepackage[numbers, sort&compress]{natbib}
\makeatletter 
\renewcommand\@biblabel[1]{#1.} 
\makeatother %

\newcommand\BibTeX{{\rmfamily B\kern-.05em \textsc{i\kern-.025em b}\kern-.08em
T\kern-.1667em\lower.7ex\hbox{E}\kern-.125emX}}

\newcommand{\bigCI}{\!\perp\!\!\!\perp}
\newcommand{\nbigCI}{\not\!\perp\!\!\!\perp}

\newcommand{\R}{R }
\newcommand{\Se}{{\rm Se}}
\newcommand{\Sp}{{\rm Sp}}
\newcommand{\See}{{\rm Se} }
\newcommand{\Spp}{{\rm Sp} }
\newcommand{\vect}[1]{\boldsymbol{\textbf{#1}}}
\newcommand{\J}{{\rm J}}
\newcommand{\JX}{{\rm J_{X}}}

\newcommand{\SeX}{{\rm Se}_{X}}
\newcommand{\SpX}{{\rm Sp}_{X}}
\newcommand{\SeZ}{ {\rm Se}_{Z_{1}} }
\newcommand{\SpZ}{ {\rm Sp}_{Z_{1}} }
\newcommand{\SeZZ}{ {\rm Se}_{Z_{2}} }
\newcommand{\SpZZ}{ {\rm Sp}_{Z_{2}} }

\newcommand{\SeIGS}{{\rm Se}_{X}^{\mathrm{IGS}}}
\newcommand{\SpIGS}{{\rm Sp}_{X}^{\mathrm{IGS}}}
\newcommand{\dSeIGS}{\Delta {\rm Se}^{\mathrm{IGS}}}
\newcommand{\dSpIGS}{\Delta {\rm Sp}^{\mathrm{IGS}}}
\newcommand{\SenIGS}{{\rm Se}_{X}^{\overline{\mathrm{IGS}}}}
\newcommand{\SpnIGS}{{\rm Sp}_{X}^{\overline{\mathrm{IGS}}}}
\newcommand{\dSenIGS}{\Delta {\rm Se}^{\overline{\mathrm{IGS}}}}
\newcommand{\dSpnIGS}{\Delta {\rm Sp}^{\overline{\mathrm{IGS}}}}

\newcommand{\ZCRSand}{Z^{\mathrm{CRS\_A}} }
\newcommand{\SeCRSand}{{\rm Se}_{X}^{\mathrm{CRS\_A}}}
\newcommand{\SpCRSand}{{\rm Sp}_{X}^{\mathrm{CRS\_A}}}
\newcommand{\dSeCRSand}{\Delta {\rm Se}^{\mathrm{CRS\_A}} }
\newcommand{\dSpCRSand}{\Delta {\rm Sp}^{\mathrm{CRS\_A}} }
\newcommand{\SenCRSand}{{\rm Se}_{X}^{\overline{\mathrm{CRS\_A}}} }
\newcommand{\SpnCRSand}{{\rm Sp}_{X}^{\overline{\mathrm{CRS\_A}}}}
\newcommand{\dSenCRSand}{\Delta {\rm Se}^{\overline{\mathrm{CRS\_A}}} }
\newcommand{\dSpnCRSand}{\Delta {\rm Sp}^{\overline{\mathrm{CRS\_A}}} }

\newcommand{\ZCRSor}{Z^{\mathrm{CRS\_O}} }
\newcommand{\SeCRSor}{{\rm Se}_{X}^{\mathrm{CRS\_O}} }
\newcommand{\SpCRSor}{{\rm Sp}_{X}^{\mathrm{CRS\_O}} }
\newcommand{\SenCRSor}{{\rm Se}_{X}^{\overline{\mathrm{CRS\_O}}}}
\newcommand{\SpnCRSor}{{\rm Sp}_{X}^{\overline{\mathrm{CRS\_O}}}}
\newcommand{\dSeCRSor}{\Delta {\rm Se}^{\mathrm{CRS\_O}}}
\newcommand{\dSpCRSor}{\Delta {\rm Sp}^{\mathrm{CRS\_O}}}
\newcommand{\dSenCRSor}{\Delta {\rm Se}^{\overline{\mathrm{CRS\_O}}}}
\newcommand{\dSpnCRSor}{\Delta {\rm Sp}^{\overline{\mathrm{CRS\_O}}}}

\newcommand{\ZDA}{Z^{\mathrm{DA}}}
\newcommand{\ZnDA}{Z^{\overline{\mathrm{DA}}}}
\newcommand{\SeDA}{{\rm Se}_{X}^{\mathrm{DA}}}
\newcommand{\SpDA}{{\rm Sp}_{X}^{\mathrm{DA}}}
\newcommand{\dSeDA}{{\Delta {\rm Se}^{\mathrm{DA}}}}
\newcommand{\dSpDA}{{\Delta {\rm Sp}^{\mathrm{DA}}}}
\newcommand{\SenDA}{{\rm Se}_{X}^{\overline{\mathrm{DA}}}}
\newcommand{\SpnDA}{{\rm Sp}_{X}^{\overline{\mathrm{DA}}}}
\newcommand{\dSenDA}{\Delta {\rm Se}^{\overline{\mathrm{DA}}}}
\newcommand{\dSpnDA}{\Delta {\rm Sp}^{\overline{\mathrm{DA}}}}

\newcommand{\dSeM}{\Delta {\rm Se}^{\mathrm{M}}}
\newcommand{\dSpM}{\Delta {\rm Sp}^{\mathrm{M}}}
\newcommand{\SenM}{{\rm Se}_{X}^{\overline{\mathrm{M}}}}
\newcommand{\SpnM}{{\rm Sp}_{X}^{\overline{\mathrm{M}}}}
\newcommand{\dSenM}{\Delta {\rm Se}^{\overline{\mathrm{M}}}}
\newcommand{\dSpnM}{\Delta {\rm Sp}^{\overline{\mathrm{M}}}}

\begin{document}

\title[The cost of not having a perfect reference]{The cost of not having a perfect reference in diagnostic accuracy studies: theoretical results and a web visualisation tool}

\author{Ana Subtil, M. Ros\'{a}rio Oliveira and Ant\'{o}nio Pacheco}

\address{CEMAT and Departamento de Matematica, Instituto Superior Tecnico, Universidade de Lisboa\\
Av. Rovisco Pais,
1049-001 LISBOA,
Portugal.\\ E-mail: anasubtil@tecnico.ulisboa.pt}

\begin{abstract}
Dichotomous diagnostic tests are widely used to detect the presence or absence of a biomedical condition of interest. A rigorous evaluation of the accuracy of a diagnostic test is critical to determine its practical value. Performance measures, such as the sensitivity and specificity of the test, should be estimated by comparison with a  gold standard. Since an error-free reference test is frequently missing, approaches based on available imperfect diagnostic tests are used, namely: comparisons with an imperfect gold standard or with a composite reference standard, discrepant analysis, and latent class models. 
			
In this work, we compare these methods using a theoretical approach based on analytical expressions for the deviations between the sensitivity and specificity according to each method, and the corresponding true values. We explore the impact on the deviations of varying conditions: tests sensitivities and specificities, prevalence of the condition and local dependence between the tests. An \R interactive graphical application is made available for the visualisation of the outcomes. Based on our findings, we discuss the methods validity and potential usefulness.
\end{abstract}


\maketitle

\section{Introduction}
\label{sec:intro}

Diagnostic test accuracy studies are crucial to determine the test's ability to discriminate between the presence or absence of a certain target condition (such as disease, infection or parasite) and thus to establish the practical value of a new diagnostic test. Two commonly used diagnostic test performance measures are the sensitivity ({\rm Se}), the probability that the test result is positive given that the subject has the target condition, and the specificity ({\rm Sp}), the probability that the test result is negative given that the subject does not have the target condition. Ideally, these measures would be estimated by comparison with a perfect reference test or gold standard (GS), an error-free diagnostic procedure with $\textrm{Se}=\textrm{Sp}=1$. A GS would determine with certainty the status of the condition of interest in the individual and thereby enable the estimation of any diagnostic test performance measures.

However, the ideal situation of having a perfect reference test is often impossible, due to budget, ethical or technical restrictions, or even because there is no GS for the target condition. Begg \cite{Begg1987} argues that more accurate tests are often more expensive and/or invasive, disadvantages which may hinder their use. In practice, it is rare for a reference standard to be completely error-free \cite{Walter2012}, as almost all tests are subject to potential error, even biopsy or autopsy \cite{Begg1987}. Therefore, the definitive diagnosis required for the correct accuracy estimation of a new diagnostic test may actually be impossible to obtain. 

The performance evaluation process that new tests undergo before being introduced into practice should be sound and rigorous to produce trustworthy conclusions \cite{Hadgu2005,Reitsma2009}. Methodological flaws in diagnostic studies can lead to biased estimates of the index test accuracy. Such erroneous findings can have a considerable impact in practice \citep{Bossuyt2003}, since diagnostic tests deliver key information for decision-making in the biomedical context.

Diverse sources of bias may affect diagnostic test accuracy studies and distort the estimated accuracy of a diagnostic test \citep{Zhou2009}. In our work we will only focus on the bias arising from evaluating the accuracy of a diagnostic test when a GS is missing. Our aim is to gain insight into alternative methods used in this context and offer guidance towards an adequate choice of test performance evaluation method. The need for further developments and methodological research regarding this topic has been emphasized by \cite{Rutjes2007} and \cite{Miller2012}. 

Authors in \cite{Walter1988,Hui1998,Rutjes2007,Collins2014} provide a broad overview on the variety of methods found in the literature to address the problem of evaluating the accuracy of a diagnostic test in the absence of a GS. We address in the present work some of the most widespread and undemanding among these methods: imperfect gold standard (IGS), composite reference standard (CRS), with the "and" rule and the "or" rule, discrepant analysis (DA) and latent class models (LCM), all of which rely on available imperfect diagnostic tests to estimate the index test accuracy.  

When a GS is missing, a straightforward approach to evaluate the performance of a new test consists of adopting as reference an imperfect test which is perceived as the best available test for the target condition. Since this test, named imperfect gold standard, is not error-free, it will potentially misclassify some subjects regarding the target condition and thus bias the estimates of the performance measures of the index test. This type of bias, called imperfect gold standard bias \citep{Baughman2008}\footnote{Other designations can be found in the literature, such as reference test bias \citep{Thibodeau1981}, imperfect reference standard bias \citep{Zhou2009} or imperfect reference bias \citep{Naakt2013}.}, has been investigated by several authors over the years \citep[e.g.,][]{Gart1966,Greenberg1969,Staquet1981,Thibodeau1981,Boyko1988,Baughman2008,Walter2012}. 

As an alternative to using a single imperfect reference, multiple imperfect tests may be combined into a composite reference standard, according to a fixed rule \citep{Alonzo1999,Baughman2008}. The rational behind this approach is that the CRS will be more accurate in terms of \See or \Spp than each component test individually. Therefore, the bias that would potentially affect the estimates of the index test accuracy measures, if a single diagnostic test was deemed as an IGS, should be partly reduced by using the CRS. Generally, there is a trade-off between \See and \Spp of the CRS when compared to the individual component tests. Indeed, the fixed rule used to combine the tests into a CRS may improve the \See but worsen the \Sp, or vice-versa. This approach has the merit of incorporating prior available information about the tests, since the rule adopted to combine them should derive from previous knowledge of the tests.  

Discrepant analysis is an approach that aims to overcome the misclassification errors due to the use of an IGS. In DA, the observations for which the new test and the IGS disagree are reassessed by a second test, called resolver test, which may even be a GS. The problems associated with DA have been largely discussed in the literature and, with rare exceptions \citep{Schachter2001}, this method has been strongly criticized and its use discouraged \citep{Hadgu1997,Miller1998,Green1998,Lipman1998,Hadgu2000,Mcadam2000,Sternberg2001}. These authors argue that, even if DA is an intuitive method, in fact it is inherently biased, overestimating test accuracy measures in most situations, and unscientific, because the new test is used to determine the true target condition status, presumably leading to incorporation bias \citep{Zhou2009}. In agreement with these claims, the US Food and Drug Administration also states that DA is an inappropriate method to estimate a diagnostic test \See and \Spp \citep{Food2003}.

Despite all the criticism, we include DA in our study, in order to rigorously evaluate its performance, since this method relies on a very intuitive and appealing strategy, which in fact is used in other fields of study. We anticipate that DA surprisingly arises as the preferable method in some scenarios.

In the context of diagnostic test studies, a widely used latent class model admits multiple binary manifest variables, that express the results of imperfect diagnostic tests, and an underlying binary latent variable, which defines two latent classes, that represent the presence or absence of a target condition. This two-latent class model is used to estimate the tests accuracy measures and the prevalence of the condition \citep{Hui1980,Albert2004,Pepe2007,Collins2014}. 

The simplest latent class models presume conditional independence between the tests results, i.e., manifest variables expressing the tests outcomes are independent for fixed values of the latent variable. The so-called Hypothesis of Conditional or local Independence (HCI) is a rather questionable assumption in many practical situations \cite{Boyko1988,Pepe2007,Reboussin2008,Walter2012}. Criticisms have been directed towards LCM regarding the difficulty in evaluating the basic assumption of conditional independence \citep{Subtil2012} and the bias resulting from the use of LCM when this assumption does not hold \citep[e.g.,][]{Vacek1985,Alonzo1999,Albert2004,Pepe2007,Collins2014}. 

LCM that relax the conditional independence assumption have been proposed in the literature, admitting either maximum likelihood estimation or bayesian inference \citep[for example,][]{Vacek1985,Qu1996,Dendukuri2001}, but are not addressed in this paper. Albert and Dodd \cite{Albert2004} have shown that LCM with different local dependence structures can adjust equally well to the data, but lead to different accuracy estimates. Identifiability problems have also been pointed out \citep{Goodman1974,Bartholomew2011}, as a consequence of the large number of parameters to estimate. 

As mentioned earlier, IGS, CRS (with the "and" rule and the "or" rule), DA and LCM are the methods addressed in this work. We derive algebraic expressions for the deviations between each method's \See and \Sp, and the corresponding true values, to evaluate and compare the methods. This theoretical approach, which removes the confounding effect of the sampling scheme, aims to provide a better understanding of the evaluation approaches under study, by clarifying the magnitude and direction of the deviations from the true values of \See and \Sp, in view of varying factors: the tests \See and \Sp, the prevalence of the target condition and the magnitude of the conditional dependence between the tests.

The theoretical unified approach we adopt to jointly investigate the aforementioned methods aims to gather comparable and consistent findings on the methods to give insight into their comparative value, and thus provide recommendations on their use in practical situations. 

An interactive graphical web application, developed using the \R Shiny package \cite{shiny}, is made available along with this article, allowing users to visualise how the investigated methods \See and \Spp theoretical deviations vary under different conditions. By making it possible to experiment distinct settings, and mimic real cases if needed, this application intends to make the theoretical findings more tangible and comprehensible, and hence useful in real diagnostic accuracy studies.  

The paper is structured as follows. We start with Section~\ref{sec:intro}, which introduces the subject being studied, as well as our approach to the problem, and gives a brief description of the accuracy evaluation methods under comparison. Section~\ref{sec:notation.setup} presents the notation and the setup adopted in our study. The following sections present the \See and \Spp analytical expressions derived for the different methods: the IGS is addressed in Section~\ref{sec:IGS}, CRS in Section~\ref{sec:CRS}, DA in Section~\ref{sec:DA}, a unified approach between the previous methods in Section~\ref{sec:IGS-CRS-DA}, and finally LCM in Section~\ref{sec:LCM}. Section~\ref{sec:Covariances_limits} reports restrictions imposed on the parameters that model local dependence.  Based on an \R Shiny application we developed to interactively visualise the outcomes of the theoretical expressions, we present a practical application in Section~\ref{sec:application}. We conclude with Section~\ref{sec:discussion}, which states final remarks and directions for future work.


\section{Notation and setup}
\label{sec:notation.setup}


\subsection{Notation}
\label{subsec2.1:notation}

Let us admit a binary variable, $Y$, whose categories $\{0,1\}$ indicate the presence or absence of a certain condition of interest (e.g., disease, infection or parasite). $Y$ takes the value 1 if the condition is present and 0 otherwise. Moreover, let us define $\eta=P(Y=1)$, the prevalence of the condition of interest, i.e., the proportion of individuals with the condition in the population.

Suppose that $\vect{X}=(X_{1},...,X_{p})^\mathsf{T}$ is a vector of $p$ binary variables that express the results of $p$ dichotomous diagnostic tests. $X_{i}$ takes the value $1$ if the $i$-th test is positive, and $0$ otherwise ($i=1,...,p$). We define \See and \Spp of the $i$-th test as 
\begin{equation}\label{Se_Sp}
\Se_{X_{i}}=P(X_{i}=1|Y=1)  \;\; \;\;   \text{and} \;\; \;\;  \Sp_{X_{i}}=P(X_{i}=0|Y=0).\\
\end{equation}

The joint probability of $X_{i}$ and $X_{j}$ is given by
\begin{eqnarray} \nonumber
P(X_{i}=x_{i},X_{j}=x_{j}) &=&\eta P(X_{i}=x_{i},X_{j}=x_{j}|Y=1) \\
&& + (1-\eta) P(X_{i}=x_{i},X_{j}=x_{j}|Y=0), \label{eq:JointPrXiXj}
\end{eqnarray} 
where $x_{i},x_{j}=0,1$, $i \neq j$, and $i,j=1,...,p$.

And the joint probability of $X_{i}$ and $X_{j}$ conditional on the category $\{Y=y\}$  emerges as 
\begin{eqnarray} \nonumber
P(X_{i}=x_{i},X_{j}=x_{j}|Y=y) &=& P(X_{i}=x_{i}|Y=y) P(X_{j}=x_{j}|Y=y)\\
&& +(-1)^{x_{i}-x_{j}}cov(X_{i},X_{j}|Y=y). \label{eq:JointPrXiXj_conditional_onY}
\end{eqnarray}

Suppose that $X_{i}$ and $X_{j}$ are independent conditional on the category of $Y$, expressed as $X_{i} \bigCI X_{j} \mid Y=y$, where $y=0,1$, $i \neq j$, and $i,j=1,...,p$. That is to say, HCI is valid. Accordingly, $cov(X_{i},X_{j}|Y=y)=0$. Hence, \eqref{eq:JointPrXiXj_conditional_onY} simplifies into 
\begin{align}\label{eq:JointPrXiXj_conditional_onYHIC}
P(X_{i}=x_{i},X_{j}=x_{j}|Y=y)=P(X_{i}=x_{i}|Y=y) P(X_{j}=x_{j}|Y=y).
\end{align}

Let us define the Youden's index \citep{Youden1950}, denoted by \J, a statistic that measures the performance of diagnostic tests, which is noticeable in some of the expressions presented further ahead in this document. For the $i$-th test, it is defined as 
\begin{align}\label{eq:Youden_index}
\J_{X_{i}} = \Se_{X_{i}} + \Sp_{X_{i}} - 1.
\end{align}

The Youden's index possible values range between -1 and 1. The index takes the value 1 in face of a perfect diagnostic test, with $\Se_{X_{i}}=\Sp_{X_{i}}=1$, and the value of -1 for a test with null \See and \Sp, $\Se_{X_{i}}=\Sp_{X_{i}}=0$. Moreover, the Youden's index has the value zero when the probability of a positive (negative) test result is the same regardless of the presence or absence of the target condition. Naturally, a test with such characteristics is useless. In fact, for a diagnostic test to be useful in the diagnostic process, it should verify $\Se_{X_{i}}+\Sp_{X_{i}} > 1$, which can also be expressed as $\J_{X_{i}}>0$. These conditions mean that it is more likely to observe a positive (negative) test result given a true positive (negative) subject than otherwise.


\subsection{Setup}
\label{subsec2.2:setup}

In this paper, we adopt an hypothetical setup to assess and compare the potential of using the aforementioned alternative methods of diagnostic test evaluation. We suppose that we want to evaluate the accuracy of a new diagnostic test for a certain condition, $X$, in the absence of a GS. We also assume that only two imperfect diagnostic tests for the same condition, $Z_1$ and $Z_2$, are available, and the true state of the condition, $Y$, cannot be observed. 

In our setup, we admit the straightforward HCI assumption, according to which $X$, $Z_1$, and $Z_2$ are conditionally independent given the true state of the target condition, $Y$:
\begin{align}
X \bigCI Z_{1} \mid Y=y, \; X \bigCI Z_{2} \mid Y=y, \; \; \textrm{ and } \; \; Z_{1} \bigCI Z_{2}\mid Y=y, \textrm{ with }y=0,1.   \label{eq:SetupIndep}
 \end{align}

Additionally, since the HCI is not a realistic assumption in many practical situations, we also study, without loss of generality,  a dependence structure which accommodates local dependence between the new test $X$ and the IGS $Z_1$, while the HCI is presumed valid for the remaining pairs of tests. To summarise, according to this dependence structure, we have
 \begin{align}
X \nbigCI Z_{1} \mid Y=y, \; X \bigCI Z_{2} \mid Y=y, \; \; \textrm{ and } \; \; Z_{1} \bigCI Z_{2}\mid Y=y, \textrm{ with }y=0,1. \label{eq:SetupDep}
 \end{align}
 
Given the conditional joint probability of the tests in \eqref{eq:JointPrXiXj_conditional_onY}, and denoting $\xi=cov(X,Z_{1}|Y=1)$ and $\epsilon=cov(X,Z_{1}|Y=0)$, under the dependence structure defined by \eqref{eq:SetupDep}, we have:
\begin{eqnarray} \label{eq:JointPrXZ1_conditional_onY}
P(X=x,Z_{1}&=z_{1}|Y=1)=P(X=x|Y=1) P(Z_{1}=z_{1}|Y=1)+(-1)^{x-z_{1}}\xi,\\
P(X=x,Z_{1}&=z_{1}|Y=0)=P(X=x|Y=0) P(Z_{1}=z_{1}|Y=0)+(-1)^{x-z_{1}}\varepsilon.  \nonumber 
\end{eqnarray}

It follows that the HCI case in \eqref{eq:SetupIndep} is a special case of \eqref{eq:JointPrXZ1_conditional_onY} with $\xi=\varepsilon=0$.

For simplicity reasons, we restricted the local dependencies in our setup to \eqref{eq:SetupDep}. However, other structures of local dependence among the diagnostic tests are possible, for which developments similar to the ones presented in this paper can be formulated. 

Much of the literature on local dependence between diagnostic tests merely investigates positive dependence, based on the claim that it is biologically more plausible than negative dependence \cite[e.g.,][]{Gardner2000,Dendukuri2001}. This type of dependence is conceivable, for instance, between diagnostic tests with a similar biological basis \cite{Vacek1985,Menten2008}, or between tests that more easily detect severe cases than mild or weak ones, leading to concordant true positive and false negative outcomes, and thus to positive dependence in the class of subjects with the disease \citep{Pepe2007,Walter2012}. Nevertheless, negative conditional dependence may also be reasonable \cite{Walter2012}, as when diagnostic tests identify different subgroups of subjects with the condition, potentially leading to discordant true positive and false negative results, i.e, local negative dependence \cite{Menten2008}. In view of this, in our setup, both positive and negative conditional dependence are anticipated.

We present theoretical expressions for the deviations between each method accuracy measures (\See and \Sp) of test $X$  and the corresponding true values.

Let $\Se_{X}$ be the true sensitivity of $X$ and $\Se_{X}^{\mathrm{M}}$ the theoretical expression for the sensitivity of $X$ determined by the method $\mathrm{M}$ under the assumption of HCI, which changes to $\Se_{X}^{\overline{\mathrm{M}}}$ under the dependence structure defined in \eqref{eq:SetupDep}.

 The deviation between $\Se_{X}^{M}$ and $\Se_{X}$ is given by
 \begin{equation}\label{eq:deltaSe}
 \Delta \Se^{\mathrm{M}}=\Se _{X}^{\mathrm{M}}-\Se_{X},
  \end{equation}
  and the equivalent expression under the violation of HCI according to \eqref{eq:SetupDep} is
   \begin{equation}\label{eq:deltaSe_nHIC}
   \Delta \Se^{\overline{\mathrm{M}}}=\Se_{X}^{\overline{\mathrm{M}}}-\Se_{X}.
   \end{equation}
   
    Similar notation is used for specificity: $\Delta \Sp^{\mathrm{M}}=\Sp_{X}^{\mathrm{M}}-\Sp_{X}, \textrm{ under HCI, and } \Delta \Sp^{\overline{\mathrm{M}}}=\Sp_{X}^{\overline{\mathrm{M}}}-\Sp_{X}, \textrm{ otherwise}$.

We also remark that the approach we undertake could be extended to include other diagnostic accuracy measures. In fact, analogous theoretical expressions to the ones obtained for \See and \Spp could be derived for measures such as positive and negative predictive value  (PPV and NPV, respectively). PPV (NPV) is defined as the probability that the subject is a true positive (negative) given that the test result is positive (negative).

Each method under scrutiny and corresponding deviations are presented and deduced in the next sections. 


\section{Imperfect gold standard}
\label{sec:IGS}

We start by the IGS, a rather straightforward method, in which an imperfect test is used as a reference. 

Suppose that $X$ is the test under study for which we aim to estimate the performance measures by comparison with an IGS. Take $Z_{1}$ as such an IGS, characterized  by 
\begin{align}
P(Z_{1}=1) = \eta \Se_{Z_{1}} +(1-\eta)\left(1-\Sp_{Z_{1}}\right).  \label{eq:PrZ1_IGS}
\end{align}


The \See and \Spp of $X$ by comparison with the IGS $Z_{1}$ are defined as
\begin{align}
\SeIGS &= P(X=1|Z_{1}=1),  \\ 
\SpIGS &= P(X=0|Z_{1}=0).
\end{align}  

$\Se_{X}^{\mathrm{IGS}}$ and $\Sp_{X}^{\mathrm{IGS}}$ can be expressed as functions of the characteristics of the tests and the prevalence. For the \Se, under the HCI, we have
\begin{align}
\SeIGS = \frac{\eta \Se_{X}\Se_{Z_{1}}+(1-\eta) (1-\Sp_{X})(1-\Sp_{Z_{1}})}{P(Z_{1}=1)}.   \label{eq:Se_IGS_HCI}
\end{align} 
 
Hence, the deviation emerges as
\begin{align}  
\dSeIGS &=-\frac{(1-\eta) (1-\Sp_{Z_{1}})(\Se_{X}+\Sp_{X}-1)}{P(Z_{1}=1)} \nonumber\\ 
	    &=-\frac{(1-\eta) (1-\Sp_{Z_{1}})\JX}{P(Z_{1}=1)}. \label{eq:DeltaSe_IGS_HCI}
\end{align}

It follows that $\dSeIGS$ is non-positive, since we assume that $\JX>0$, or else $X$ would be a useless test. Given that both $\JX=0$ and $\eta=1$ are unreasonable conditions, $\dSeIGS$ is null only when the IGS $Z_{1}$ has perfect specificity, i.e., $\Sp_{Z_{1}}=1$. Besides this particular case, in which unbiased \See estimates could be obtained, in most practical situations, using an IGS under HCI would lead to \See underestimation.


When the HCI is violated according to the dependence structure \eqref{eq:SetupDep},
\begin{align}
\SenIGS = \SeIGS +\frac{\eta \xi+(1-\eta) \varepsilon}{P(Z_{1}=1)} \label{eq:Se_IGS_nHCI}
\end{align}
and the $\Se$ deviation is given by
\begin{align} \label{eq:deltaSe_IGS_nHCI}
\dSenIGS =\dSeIGS + \frac{\eta \xi+(1-\eta) \varepsilon}{P(Z_{1}=1)}. 
\end{align}

These two expressions, \eqref{eq:Se_IGS_nHCI} and \eqref{eq:deltaSe_IGS_nHCI}, are generalizations of the basic HCI ones, \eqref{eq:Se_IGS_HCI} and \eqref{eq:DeltaSe_IGS_HCI}, respectively, in which a term expressing the conditional dependence is added to the HCI expressions, assuming positive or negative values depending on $\xi$ and $\varepsilon$.


Regarding the specificity, in the simpler HCI case, we have
\begin{align}
\SpIGS= \frac{\eta (1-\Se_{X})(1-\Se_{Z_{1}})+(1-\eta) \Sp_{X} \Sp_{Z_{1}}}{P(Z_{1}=0)}, \label{eq:Sp_IGS_HCI}
\end{align}
where
\begin{align}
P(Z_{1}=0)=1-P(Z_{1}=1)
\label{eq:PrZ0_IGS}
\end{align}
\\
Hence, the deviation is
\begin{align} \label{eq:DeltaSp_IGS_HCI}
\dSpIGS &=-\frac{\eta (1-\Se_{Z_{1}}) (\Se_{X}+\Sp_{X}-1)}{P(Z_{1}=0)} \\ \nonumber
                 &=-\frac{\eta (1-\Se_{Z_{1}}) \JX}{P(Z_{1}=0)}.
\end{align}

Accordingly, $\dSpIGS$ is also non-positive, assuming $\JX>0$. The null deviation occurs when the IGS, $Z_{1}$, has perfect \Se, that is, $\Se_{Z_{1}}=1$. Additionally, $\dSpIGS=0$ if $\JX=0$ or if $\eta=0$, which are both unlikely, since the former would indicate a worthless index test, and the latter a null prevalence. 


In the case of HCI violation, the specificty emerges as
\begin{align}
\SpnIGS= \SpIGS+\frac{\eta \xi+(1-\eta) \varepsilon}{P(Z_{1}=0)}, \label{eq:Sp_IGS_nHCI}
\end{align}
\\
with the corresponding deviation as
\begin{align}
\dSpnIGS= \dSpIGS+\frac{\eta \xi + (1-\eta) \varepsilon}{P(Z_{1}=0)}.      \label{eq:DeltaSp_IGS_nHCI}
\end{align}

It follows from \eqref{eq:deltaSe_IGS_nHCI} and \eqref{eq:DeltaSp_IGS_nHCI} that the \See and \Spp deviations under the HCI violation are composed of two terms: a first one reflecting the method's imperfection, shared with the HCI case, and a second term quantifying the local dependence effects. The first term should contribute to \Se/\Spp underestimation, except for a few particular cases of unbiased estimation, while the second could lead either to under or overestimation, depending on the local dependence. Hence the two effects contribute in opposite directions when the local dependence term is positive. 


\section{Composite Reference Standard}
\label{sec:CRS}

The chosen reference standard is crucial for obtaining reliable estimates of the index test accuracy from its comparison against the reference. We will now address the use as a reference of a CRS obtained from the combination of two imperfect tests according to a fixed rule. Previous information available on the tests should support the choice of rule to combine the tests. We will address two rules that can be used to combine two diagnostic tests: the "and" rule, in which a positive outcome emerges if both component tests indicate a positive result, and the "or" rule, in which a positive result occurs if any of the two tests is positive. 

The bias of the index test accuracy estimates associated with the use of a CRS under the "or" rule has been addressed in works such as \cite{Alonzo1999}, \cite{Baughman2008} and \cite{Schiller2015}, the latter of which also studies the CRS with the "and" rule.

Besides the combination of imperfect tests to form a CRS that we focus on this article, others combinations can be found in the literature. As an example, \cite{Hadgu2012} assessed a different method called PISA, used in recent years to evaluate nucleid acid amplification tests for detecting Chlamydia tractomatis and Neisseria gonorrhea. 

\subsection{Composite Reference Standard - ``and" rule}
\label{subsec4.1:CRS-and}

Let $\ZCRSand$ be a binary variable that expresses the results of combining two diagnostic tests according to the "and" rule. Hence $\ZCRSand$ is positive if both tests are positive, and negative otherwise, i.e., if any of the tests is negative,
\begin{equation}
\ZCRSand = 
\begin{cases}  \label{eq:Z_CRS_AND}
1,   & \text{if } (Z_{1}=1 \wedge Z_{2}=1) \\
0,   & \text{if } (Z_{1}=0 \vee Z_{2}=0).
\end{cases} 
\end{equation} 
\\
Accordingly, we can define the probabilities
\begin{align} 
\label{eq:PrZ_CRS}
P(\ZCRSand=1) = \eta \Se_{Z_{1}} \Se_{Z_{2}}+(1-\eta)(1-\Sp_{Z_{1}})(1-\Sp_{Z_{2}}) \; \; \text{and} \\  
P(\ZCRSand=0) = 1-P(\ZCRSand=1), \nonumber
\end{align}
which remain unchanged whether under the HCI or the dependence structure in \eqref{eq:SetupDep}, because $Z_{1}$ and $Z_{2}$ are conditionally independent in both cases.

Under the HCI, the sensitivity arises as 
\begin{align}
\SeCRSand = \frac{\eta \Se_{X} \Se_{Z_{1}} \Se_{Z_{2}}+(1-\eta) (1-\Sp_{X}) (1-\Sp_{Z_{1}}) (1-\Sp_{Z_{2}})}{P(\ZCRSand=1)},  \label{eq:Se_CRS_AND_HCI}
\end{align}
and the sensitivity deviation as
\begin{align}
\dSeCRSand &= -\frac{(1-\eta) (1-\Sp_{Z_{1}}) (1-\Sp_{Z_{2}}) (\Se_{X}+\Sp_{X}-1)}{P(\ZCRSand=1)} \label{eq:DeltaSe_CRS_AND_HCI}\\
 \notag                     &= -\frac{(1-\eta) (1-\Sp_{Z_{1}}) (1-\Sp_{Z_{2}}) \JX}{P(\ZCRSand=1)}.
\end{align}
\\
Admitting the dependence structure described in \eqref{eq:SetupDep}, where $X \nbigCI Z_{1} \mid Y=y$, with $y=0,1$, we derive
\begin{align}
\SenCRSand = \SeCRSand+\frac{\eta  \Se_{Z_{2}} \xi +(1-\eta)  (1-\Sp_{Z_{2}}) \varepsilon}{P(\ZCRSand=1)}.  \label{eq:Se_CRS_AND_nHCI}
\end{align}
\\
The $\Se$ deviation emerges as
\begin{align}
\dSenCRSand =\dSeCRSand+\frac{\eta  \Se_{Z_{2}} \xi +(1-\eta)  (1-\Sp_{Z_{2}}) \varepsilon}{P(\ZCRSand=1)}. \label{eq:DeltaSe_CRS_AND_nHCI}
\end{align}
\\
As for the specificity, under the HCI, 
\begin{align}
\SpCRSand= \frac{\eta (1-\Se_{X}) (1-\Se_{Z_{1}} \Se_{Z_{2}})+(1-\eta) \Sp_{X} (\Sp_{Z_{1}}+\Sp_{Z_{2}}-\Sp_{Z_{1}} \Sp_{Z_{2}})}{P(\ZCRSand=0)},  \label{eq:Sp_CRS_AND_HCI}
\end{align}
\\
hence, the specificity deviation emerges as
\begin{align}
\dSpCRSand &= -\frac{\eta  (1-\Se_{Z_{1}} \Se_{Z_{2}})(\Se_{X}+\Sp_{X}-1) }{P(\ZCRSand=0)} \label{eq:DeltaSp_CRS_AND_HCI} \\
\notag &= -\frac{\eta  (1-\Se_{Z_{1}} \Se_{Z_{2}}) \JX }{P(\ZCRSand=0)}.
\end{align}
\\
In the case of local dependence as defined by \eqref{eq:SetupDep},
\begin{align}
\SpnCRSand = \SpCRSand+\frac{\eta  \Se_{Z_{2}} \xi +(1-\eta)  (1-\Sp_{Z_{2}}) \varepsilon}{P(\ZCRSand=0)},  \label{eq:Sp_CRS_AND_nHCI}
\end{align}
\\
whereby the \Spp deviation becomes
\begin{align}
\dSpnCRSand  =\dSpCRSand+\frac{\eta  \Se_{Z_{2}} \xi +(1-\eta)  (1-\Sp_{Z_{2}}) \varepsilon}{P(\ZCRSand=0)}. \label{eq:DeltaSp_CRS_AND_nHCI}
\end{align}

Just as described earlier for IGS, the deviations $\dSenCRSand$ and $\dSpnCRSand$, in  \eqref{eq:DeltaSe_CRS_AND_nHCI} and \eqref{eq:DeltaSp_CRS_AND_nHCI}, have two components: one reflecting the method's imperfection, shared with the HCI case, and a second one expressing the conditional dependence. The first component is non-positive (since it would be unreasonable to conceive $\JX<0$), contributing to the \See and \Spp underestimation (except for a few cases of null bias). The second component, which is due to local dependence, may be either negative or positive, depending on the conditional covariances, $\xi$ and $\varepsilon$. Hence, the local dependence effect may contribute either to reinforce or cancel the effect of the $\mathrm{CRS\_A}$ imperfection.

\subsection{Composite Reference Standard - ``or" rule}
\label{subsec4.2:CRS-or}

We admit that $\ZCRSor$ is a binary variable that expresses the results of combining two diagnostic tests, $Z_{1}$ and $Z_{2}$, using the ``or" rule, which means that $\ZCRSor$ is positive if any of the tests, $Z_{1}$ or $Z_{2}$, is positive, and $\ZCRSor$ is negative otherwise, i.e., if both tests are negative. It follows that
\begin{equation}
\ZCRSor = 
\begin{cases}
1,   & \text{if } (Z_{1}=1 \vee Z_{2}=1) \\
0,  & \text{if } (Z_{1}=0 \wedge Z_{2}=0),
\label{eq:Z_CRS_OR}
\end{cases} 
\end{equation}
and consequently
\begin{align}
\label{eq:PrZ1_CRS_OR}
P(\ZCRSor=1) = \eta (\Se_{Z_{1}} +\Se_{Z_{2}}-\Se_{Z_{1}} \Se_{Z_{2}})+(1-\eta)(1-{\rm Sp}_{Z_{1}}{\rm Sp}_{Z_{2}}) \; \; \textrm{ and }  \\
P(\ZCRSor=0)=1- P(\ZCRSor=1). \nonumber 
\end{align}
Assuming that the HCI is valid,
\begin{align}
\SeCRSor = \frac{\eta \Se_{X} (\Se_{Z_{1}}+\Se_{Z_{2}}-\Se_{Z_{1}} \Se_{Z_{2}})+(1-\eta) (1-\Sp_{X}) (1-\Sp_{Z_{1}} \Sp_{Z_{2}})}{P(\ZCRSor=1)},  \label{eq:Se_CRS_OR_HCI}
\end{align}
and the $\Se$ deviation is given by
\begin{align}
\dSeCRSor &= -\frac{(1-\eta)  (1-\Sp_{Z_{1}} \Sp_{Z_{2}}) (\Se_{X}+\Sp_{X}-1)}{P(\ZCRSor=1)}. \label{eq:DeltaSe_CRS_OR_HCI} \\
\notag &=-\frac{(1-\eta) (1-\Sp_{Z_{1}} \Sp_{Z_{2}}) \JX}{P(\ZCRSor=1)}. 
\end{align}
\\
Admitting the dependence structure in \eqref{eq:SetupDep} leads to 
\begin{align}
\SenCRSor = \SeCRSor+\frac{\eta(1-\Se_{Z_{2}})  \xi  +(1-\eta)  \Sp_{Z_{2}} \varepsilon}{P(\ZCRSor=1)},  \label{eq:Se_CRS_OR_nHCI}
\end{align}
\\
and the corresponding $\Se$ deviation is
\begin{align}
\dSenCRSor =\dSeCRSor+\frac{\eta (1-\Se_{Z_{2}})  \xi+(1-\eta) \Sp_{Z_{2}} \varepsilon}{P(\ZCRSor=1)}. \label{eq:DeltaSe_CRS_OR_nHCI}
\end{align}
\\
Regarding the \Sp, assuming the HCI, we have
\begin{align}
\SpCRSor= \frac{\eta (1-\Se_{X}) (1-\Se_{Z_{1}}) (1-\Se_{Z_{2}})+(1-\eta) \Sp_{X} \Sp_{Z_{1}} \Sp_{Z_{2}}}{P(\ZCRSor=0)},  \label{eq:Sp_CRS_OR_HCI}
\end{align}
and the specificity deviation
\begin{align}
\dSpCRSor&= -\frac{\eta(1-\Se_{Z_{1}})(1-\Se_{Z_{2}})(\Se_{X}+\Sp_{X}-1)}{P(\ZCRSor=0)} \label{eq:DeltaSp_CRS_OR_HCI} \\
\notag &=-\frac{\eta(1-\Se_{Z_{1}})(1-\Se_{Z_{2}})\JX}{P(\ZCRSor=0)}.
\end{align}
\\
As for the $\Sp$ under the dependence structure presumed in \eqref{eq:SetupDep}, 
\begin{align}
\SpnCRSor = \SpCRSor+\frac{\eta (1-\Se_{Z_{2}}) \xi  +(1-\eta) \Sp_{Z_{2}} \varepsilon}{P(\ZCRSor=0)},  \label{eq:Sp_CRS_OR_nHCI}
\end{align}
\\
and the $\Sp$ deviation arises as
\begin{align}
\dSpnCRSor =\dSpCRSor+\frac{\eta (1-\Se_{Z_{2}}) \xi  +(1-\eta) \Sp_{Z_{2}}  \varepsilon}{P(\ZCRSor=0)}. \label{eq:DeltaSp_CRS_OR_nHCI}
\end{align}

Once again the deviations under local dependence, $\dSenCRSor$ and $\dSpnCRSor$ in \eqref{eq:DeltaSe_CRS_OR_nHCI} and \eqref{eq:DeltaSp_CRS_OR_nHCI}, comprise two terms, one corresponding to the method's imperfection and one to local dependence. The term expressing the effect of $\mathrm{CRS\_O}$ leads to underestimation of either \See or \Spp (except for the few cases of null bias), while the effect of local dependence may be either negative or positive, whereby the two effects may reinforce or cancel each other.


\section{Discrepant analysis}
\label{sec:DA}

In discrepant analysis, the test whose performance we aim to evaluate, $X$, is initially compared with an imperfect diagnostic test, $Z_{1}$. If both tests agree, the corresponding result is assumed correct, but if the two tests disagree, another test, called arbiter or resolver test, $Z_{2}$, is performed to resolve the discrepancy. The result of applying DA is expressed by the binary variable $\ZDA$:
\begin{equation}
\ZDA= 
\begin{cases}
1    \text{   if   } & (X=1 \wedge Z_{1}=1)  \vee \\
                         & (X=1 \wedge Z_{1}=0 \wedge Z_{2}=1 ) \vee  (X=0 \wedge Z_{1}=1 \wedge Z_{2}=1 ) \\
0    \text{   if   } & (X=0 \wedge Z_{1}=0)  \vee \\
                         &  (X=0 \wedge Z_{1}=1 \wedge Z_{2}=0 ) \vee  (X=1 \wedge Z_{1}=0 \wedge Z_{2}=0 ).
\label{eq:Z_DA_HCI}
\end{cases} 
\end{equation}

Under the HCI, we have the following joint probabilities
\begin{align*}
& P^{\mathrm{HCI}}(X=1,Z_{1}=1)=\eta \Se_{X}\Se_{Z_{1}}+(1-\eta)(1-\Sp_{X})(1-\Sp_{Z_{1}}), \\
& P^{\mathrm{HCI}}(X=1,Z_{1}=0,Z_{2}=1)=\eta \Se_{X}(1-\Se_{Z_{1}}) \Se_{Z_{2}}+(1-\eta) (1-\Sp_{X}) \Sp_{Z_{1}}(1-\Sp_{Z_{2}}), \\
& P^{\mathrm{HCI}}(X=0,Z_{1}=1,Z_{2}=1)=\eta (1-\Se_{X}) \Se_{Z_{1}} \Se_{Z_{2}}+(1-\eta) \Sp_{X} (1-\Sp_{Z_{1}}) (1-\Sp_{Z_{2}}),
\end{align*}
which add up to obtain
\begin{align*}
P(\ZDA=1)=P^{\mathrm{HCI}}(X=1,Z_{1}=1)+P^{\mathrm{HCI}}(X=1,Z_{1}=0,Z_{2}=1)+P^{\mathrm{HCI}}(X=0,Z_{1}=1,Z_{2}=1).
\end{align*}
The \See according to DA is given by
\begin{align}
\SeDA = \frac{P^{\mathrm{HCI}}(X=1,Z_{1}=1)+P^{\mathrm{HCI}}(X=1,Z_{1}=0,Z_{2}=1)}{P(\ZDA=1)}, 	\label{eq:Se_DA_HCI}
\end{align}
and the sensitivity deviation is
\begin{align}
\dSeDA = & \frac{(1-\Se_{X})\Big[P^{\mathrm{HCI}}(X=1,Z_{1}=1)+P^{\mathrm{HCI}}(X=1,Z_{1}=0,Z_{2}=1)\Big]}{P(\ZDA=1)} \label{eq:DeltaSe_DA_HCI}\\
 & -\frac{\Se_{X}P^{\mathrm{HCI}}(X=0,Z_{1}=1,Z_{2}=1)}{P(\ZDA=1)}. \nonumber
\end{align}

Under the HCI violation in Eq. \eqref{eq:SetupDep}, we have 
\begin{align*}
& P^{\overline{\mathrm{HCI}}}(X=1,Z_{1}=1)=P^{\mathrm{HCI}}(X=1,Z_{1}=1) + \eta \xi +(1-\eta) \varepsilon,\\
& P^{\overline{\mathrm{HCI}}}(X=1,Z_{1}=0,Z_{2}=1)=P^{\mathrm{HCI}}(X=1,Z_{1}=0,Z_{2}=1)-\eta  \Se_{Z_{2}} \xi-(1-\eta) (1-\Sp_{Z_{2}}) \varepsilon,\\
& P^{\overline{\mathrm{HCI}}}(X=0,Z_{1}=1,Z_{2}=1)=P^{\mathrm{HCI}}(X=0,Z_{1}=1,Z_{2}=1)-\eta \Se_{Z_{2}}  \xi - (1-\eta)  (1-\Sp_{Z_{2}} ) \varepsilon,
\end{align*}
and then 
\begin{align*}
P(\ZnDA=1)=P(\ZDA=1)+\eta(1-2\SeZZ)\xi-(1-\eta)(1-2\SpZZ)\varepsilon.
\end{align*}

Accordingly, we obtain
\begin{align}
\SenDA= \frac{P^{\overline{\mathrm{HCI}}}(X=1,Z_{1}=1)+P^{\overline{\mathrm{HCI}}}(X=1,Z_{1}=0,Z_{2}=1)}{P(\ZnDA=1)},  \label{eq:Se_DA_nHCI}
\end{align}

and the sensitivity deviation $\dSenDA$ emerges as
\begin{align}
\dSenDA = & \frac{(1-\Se_{X})\Big[P^{\overline{\mathrm{HCI}}}(X=1,Z_{1}=1)+P^{\overline{\mathrm{HCI}}}(X=1,Z_{1}=0,Z_{2}=1)\Big]}{P(\ZnDA=1)}  \label{eq:DeltaSe_DA_nHCI} \\
& -\frac{\Se_{X}P^{\overline{\mathrm{HCI}}}(X=0,Z_{1}=1,Z_{2}=1)}{P(\ZnDA=1)}. \nonumber
\end{align}

In the case of specificity, under the HCI, we obtain the expression 

\begin{align}
\SpDA= \frac{P^{\mathrm{HCI}}(X=0,Z_{1}=0)+P^{\mathrm{HCI}}(X=0,Z_{1}=1,Z_{2}=0)}{P(\ZDA=0)}, \label{eq:Sp_DA_HCI}
\end{align}
given that
\begin{align*}
& P^{\mathrm{HCI}}(X=0,Z_{1}=0)=\eta (1-\Se_{X}) (1-Se_{Z_{1}})+(1-\eta) \Sp_{X} \Sp_{Z_{1}}, \\
& P^{\mathrm{HCI}}(X=0,Z_{1}=1,Z_{2}=0)=\eta (1-\Se(X))\Se_{Z_{1}}(1-\Se_{Z_{2}})+(1-\eta) \Sp_{X} (1-\Sp_{Z_{1}})\Sp_{Z_{2}},\\
& P^{\mathrm{HCI}}(X=1,Z_{1}=0,Z_{2}=0)=\eta \Se_{X} (1- \Se_{Z_{1}})(1- \Se_{Z_{2}})+(1-\eta) (1-\Sp_{X}) \Sp_{Z_{1}} \Sp_{Z_{2}},
\end{align*}
and also
\begin{align}
& P(Z^{\mathrm{DA}}=0)=1-P(Z^{\mathrm{DA}}=1).
\end{align}
\\
Therefore, the specificity deviation is
\begin{align}
\dSpDA   = & \frac{(1-\Sp_{X})\Big[P^{\mathrm{HCI}}(X=0,Z_{1}=0)+P^{\mathrm{HCI}}(X=0,Z_{1}=1,Z_{2}=0)\Big]}{P(\ZDA=0)} \label{eq:DeltaSp_DA_HCI}\\
\notag &-\frac{\Sp_{X}P^{\mathrm{HCI}}(X=1,Z_{1}=0,Z_{2}=0)}{P(\ZDA=0)}. 
\end{align}
\\
Admitting the local dependence described in \eqref{eq:SetupDep}, we then have the following probabilities 
\begin{align*}
& P^{\overline{\mathrm{HCI}}}(X=0,Z_{1}=0)=P^{\mathrm{HCI}}(X=0,Z_{1}=0)+ \eta \xi + (1-\eta) \varepsilon, \\
& P^{\overline{\mathrm{HCI}}}(X=0,Z_{1}=1,Z_{2}=0)=P^{\mathrm{HCI}}(X=0,Z_{1}=1,Z_{2}=0)- \eta (1-\Se_{Z_{2}}) \xi -(1-\eta)  \Sp_{Z_{2}} \varepsilon,\\
& P^{\overline{\mathrm{HCI}}}(X=1,Z_{1}=0,Z_{2}=0)=P^{\mathrm{HCI}}(X=1,Z_{1}=0,Z_{2}=0)-\eta (1-\Se_{Z_{2}}) \xi -(1-\eta) \Sp_{Z_{2}}\varepsilon,
\end{align*}
and also
\begin{align}
P(\ZnDA=0)=1-P(\ZnDA=1)
\end{align}
\\
Hence $\SpnDA$ emerges as
\begin{align}
\SpnDA = \frac{P^{\overline{\mathrm{HCI}}}(X=0,Z_{1}=0)+P^{\overline{\mathrm{HCI}}}(X=0,Z_{1}=1,Z_{2}=0)}{P(\ZnDA=0)}. \label{eq:Sp_DA_nHCI} 
\end{align}

Additionally, $\dSpnDA$, is given by:
\begin{align}
\dSpnDA   = & \frac{(1-\Sp_{X})\Big[P^{\overline{\mathrm{HCI}}}(X=0,Z_{1}=0)+P^{\overline{\mathrm{HCI}}}(X=0,Z_{1}=1,Z_{2}=0)\Big]}{P(\ZnDA=0)} \label{eq:DeltaSp_DA_nHCI}\\
\notag &-\frac{\Sp_{X}P^{\overline{\mathrm{HCI}}}(X=1,Z_{1}=0,Z_{2}=0)}{P(\ZnDA=0)}. 
\end{align}

Unlike IGS, $\mathrm{CRS\_A}$ or $\mathrm{CRS\_O}$, in the case of DA, $\SenDA$, $\SpnDA$, and the corresponding deviations, $\dSenDA$ and $\dSpnDA$, do not comprise separate terms for the method's imperfection and for the dependence effect. As a matter of fact, the conditional covariances, $\xi$ and $\varepsilon$, are not confined to a detached term in the formulas, but appear both in the numerator and denominators of \eqref{eq:Se_DA_nHCI}, \eqref{eq:DeltaSe_DA_nHCI}, \eqref{eq:Sp_DA_nHCI}, and \eqref{eq:DeltaSp_DA_nHCI}, combined with the expressions present in the formulas derived for the DA under HCI.

\section{A unified approach for IGS, CRS\_A, CRS\_O, and DA}
\label{sec:IGS-CRS-DA}

Methods $\mathrm{CRS\_A}$, $\mathrm{CRS\_O}$, and $\mathrm{DA}$ can be seen as special cases of IGS. Indeed, it is possible to express each method $\mathrm{M} \in \{\mathrm{CRS\_A},\mathrm{CRS\_O},\mathrm{DA}\}$ by means of an IGS, represented by  $\widetilde{Z}^{\mathrm{M}}$ in Table \ref{table:M_IGS}. The reformulated conditional covariances, $\widetilde{\xi}^{\mathrm{M}}=cov(X,\widetilde{Z}^{\mathrm{M}}|Y=1)$ and $\widetilde{\varepsilon}^{\mathrm{M}}=cov(X,\widetilde{Z}^{\mathrm{M}}|Y=0)$ are included in the same table. Approaching each method M as an IGS, the derivations developed in Sections \ref{sec:CRS} and \ref{sec:DA}, may be simpler if the $\Se _{X}^{\mathrm{M}}$, $\Sp _{X}^{\mathrm{M}}$, $\Delta \Se^{\mathrm{M}}$, and $\Delta \Sp^{\mathrm{M}}$, as well as the corresponding expressions for the HCI violation, are derived as special cases of \eqref{eq:Se_IGS_HCI} through \eqref{eq:DeltaSp_IGS_nHCI}. Moreover, the unification of this group of methods covered in our work, all of which rely on comparisons with imperfect references, represents a unified way of studying their deviations.  

Furthermore, this approach makes clear, by the expression obtained for $\widetilde{Z}^{\mathrm{DA}}$, that, in the case of DA, the index test $X$ is used to define the reference against which $X$ itself is evaluated. This participation of the test under evaluation in the construction of the reference has been pointed as one of the major drawbacks of DA \citep[e.g.,][]{Miller1998,Hadgu2000}. It is also due to the inclusion of $X$ in $\widetilde{Z}^{\mathrm{DA}}$ that, unlike the other methods in Table \ref{table:M_IGS}, it is not possible to express any of the deviations, $\dSenDA$ or $\dSpnDA$, in \eqref{eq:DeltaSe_DA_nHCI} and \eqref{eq:DeltaSp_DA_nHCI}, as the sum of two separate terms, one for the method's imperfection and one for the dependence effect.    

\begin{table}[]
\centering
\caption{Correspondence between the methods $\mathrm{M} \in \{\mathrm{CRS\_A},\mathrm{CRS\_O},\mathrm{DA}\}$ covered in our work and the IGS.}
\label{table:M_IGS}
\begin{tabular}{|cccc|}
\hline
M                 & $\widetilde{Z}^{\mathrm{M}}$                       & $\widetilde{\xi}^{\mathrm{M}}$ & $\widetilde{\varepsilon}^{\mathrm{M}}$ \\ \hline
IGS               & $Z_{1}$                                                & $\xi$    & $\varepsilon$        \\
$\mathrm{CRS\_A}$ & $Z_{1}Z_{2}$                                           & $-\xi(1-\SeZZ)$             & $-\varepsilon\SpZZ$             \\
$\mathrm{CRS\_O}$ & $1-(1-Z_{1})(1-Z_{2})$                                 & $\xi(1-\SeZZ)$    & $\varepsilon \SpZZ$       \\
DA                & $X Z_{1}+X(1-Z_{1})Z_{2}+(1-X)Z_{1}Z_{2}$ &          $\widetilde{\xi}^{\mathrm{DA}}$ \textsuperscript{(1)}        &      $\widetilde{\varepsilon}^{\mathrm{DA}}$ \textsuperscript{(1)}                     \\ \hline
\end{tabular}
\vspace{1ex}

     \raggedright{\textsuperscript{(1)} Lengthy expressions omitted from the table. Vide \eqref{cova_DA} and \eqref{covb_DA}, respectively, for   $\widetilde{\xi}^{\mathrm{DA}}$ and $\widetilde{\varepsilon}^{\mathrm{DA}}$.
     \begin{align}
\widetilde{\xi}^{\mathrm{DA}} & =  \SeX (1-\SeX) \Big[ \SeZ +\SeZZ(1-2 \SeZ)\Big] + \xi (1-\SeX-\SeZZ+2 \SeX \SeZZ) \label{cova_DA}.\\
\widetilde{\varepsilon}^{\mathrm{DA}} & =\SpX (1-\SpX) \Big[ \SpZ +\SpZZ(1-2 \SpZ)\Big] + \varepsilon (1-\SpX-\SpZZ+2 \SpX \SpZZ). \label{covb_DA}
\end{align}
     }
\end{table}

\begin{table}[]
\centering
\caption{Summary of the findings on the \See deviations for the methods IGS, $\mathrm{CRS\_A}$, and $\mathrm{CRS\_O}$.}
\label{table:SeDeviations}
{\def\arraystretch{1.9}
\resizebox{\columnwidth}{!}{
\begin{tabular}{|c| c c c| c c|}
\hline
                  & \multicolumn{3}{c|}{HCI} & \multicolumn{2}{c|}{HCI violation} \\ \cline{2-6} 
                  Methods  &     $\dSeM \leq 0$       &   $\dSeM = 0$   &    
Monotony: $\dSeM \searrow$ if \textsuperscript{(2) (3)}  
  &   $\dSenM\leq\dSeM$                   &   $\dSenM \geq 0$       \\  \hline
IGS               & Always     & 


  \begin{tabular}{c} 
    \begin{minipage}[t]{0.2\textwidth} \centering
     $\Sp_{Z_{1}}=1$ \\ $\vee$ \\ $\JX=0 \vee  \eta=1$ \textsuperscript{(4)}
  \end{minipage} 
  \end{tabular} 

  &

  \begin{tabular}{c} 
    \begin{minipage}[t]{0.2\textwidth} \centering
    \begin{itemize}
    \item $\Se_{X} \nearrow$
    \item $\Sp_{X} \nearrow$
    \item $\Se_{Z_{1}}\searrow$
    \item $\Sp_{Z_{1}}\searrow$
    \item $\eta \searrow$
    \end{itemize}
  \end{minipage} 
  \end{tabular}   
  &    $\eta \xi+(1-\eta)\varepsilon \leq 0$       &   \begin{tabular}{c} \begin{minipage}[t]{0.3\textwidth} \centering $\eta \xi+(1-\eta)\varepsilon$ \\ $\geq$ \\ $(1-\eta) (1-\Sp_{Z_{1}})\JX$ \end{minipage} \end{tabular}                 \\
                 &            &      &     &           &                    \\
$\mathrm{CRS\_A}$ & Always     & 

  \begin{tabular}{l} 
    \begin{minipage}[t]{0.2\textwidth} \centering
     $\Sp_{Z_{1}}=1 \vee \Sp_{Z_{2}}=1$\\ $\vee$ \\ $\JX=0 \vee  \eta=1$ \textsuperscript{(4)}
  \end{minipage} 
  \end{tabular} 
  &  \begin{tabular}{l} \begin{minipage}[t]{0.2\textwidth}
    \begin{itemize}
    \item $\Se_{X} \nearrow$
    \item $\Sp_{X} \nearrow$
    \item $\Se_{Z_{1}}\searrow$
    \item $\Sp_{Z_{1}}\searrow$ 
    \item $\Se_{Z_{2}}\searrow$
    \item $\Sp_{Z_{2}}\searrow$
    \item $\eta \searrow$
    \end{itemize}
  \end{minipage} \end{tabular}    &     $\eta \Se_{Z_{2}} \xi+(1-\eta) (1-\Sp_{Z_{2}}) \varepsilon\leq0$        &    \begin{tabular}{c} \begin{minipage}[t]{0.3\textwidth} \centering $\eta \Se_{Z_{2}} \xi+(1-\eta) (1-\Sp_{Z_{2}}) \varepsilon$ \\ $\geq$ \\ $(1-\eta) (1-\Sp_{Z_{1}})  (1-\Sp_{Z_{2}})\JX$ \end{minipage} \end{tabular}          \\
                  &            &      &                &     &               \\
$\mathrm{CRS\_O}$ & Always     & 

  \begin{tabular}{l} 
    \begin{minipage}[t]{0.2\textwidth} \centering
     $\Sp_{Z_{1}}=1 \wedge \Sp_{Z_{2}}=1$\\ $\vee$ \\ $\JX=0 \vee  \eta=1$ \textsuperscript{(4)}
  \end{minipage} 
  \end{tabular} 



&  \begin{tabular}{l} \begin{minipage}[p]{0.20\textwidth}
    \begin{itemize}
    \item $\Se_{X} \nearrow$ 
    \item $\Sp_{X} \nearrow$
    \item $\Se_{Z_{1}}\searrow$
    \item $\Sp_{Z_{1}}\searrow$ 
    \item $\Se_{Z_{2}}\searrow$
    \item $\Sp_{Z_{2}}\searrow$
    \item $\eta \searrow$
    \end{itemize}
  \end{minipage} \end{tabular}   &       $\eta (1-\Se_{Z_{2}}) \xi  +(1-\eta) \Sp_{Z_{2}} \varepsilon \leq 0$         &       \begin{tabular}{c} \begin{minipage}[p]{0.3\textwidth} \centering $\eta (1-\Se_{Z_{2}})  \xi+(1-\eta) \Sp_{Z_{2}} \varepsilon$ \\ $\geq$ \\ $(1-\eta) (1-\Sp_{Z_{1}} \Sp_{Z_{2}}) \JX$ \end{minipage} \end{tabular}          \\
                &            &      &     &                &               \\ \hline
\end{tabular}
}
}

\vspace{1ex}

\raggedright{\textsuperscript{(2)}} The symbol $\searrow$ refers to ``decreasing" and $\nearrow$ to ``increasing". \\
\raggedright{\textsuperscript{(3)}} Any of the listed conditions contributes to decrease $\dSeM$ ($\dSeM \searrow$).\\
\raggedright{\textsuperscript{(4)}} $\JX=0$ and $\eta=1$ are unreasonable conditions, since $\JX=0$ implies a valueless index test, and $\eta=1$, i.e., a prevalence of one, means that every subject in the population has the target condition. In fact, we assume $\JX>0$ and $\eta \in ]0,1[$. 

\end{table}

\begin{table}[]
\centering
\caption{Summary of the findings on the \Spp deviations for the methods IGS, $\mathrm{CRS\_A}$, and $\mathrm{CRS\_O}$.}
\label{table:SpDeviations}
{\def\arraystretch{1.9}
\resizebox{\columnwidth}{!}{
\begin{tabular}{|c|c c c|c c|}
\hline
                  & \multicolumn{3}{c|}{HCI} & \multicolumn{2}{c|}{HCI violation} \\ \cline{2-6} 
                  Methods  &     $\dSpM \leq 0$       &   $\dSpM = 0$   &    
Monotony: $\dSpM \searrow$ if \textsuperscript{(5)}  
  &   $\dSpnM\leq\dSpM$                   &   $\dSpnM \geq 0$       \\  \hline
IGS               & Always     &  

\begin{tabular}{c} 
    \begin{minipage}[t]{0.2\textwidth} \centering
     $\Se_{Z_{1}}=1$ \\ $\vee$ \\ $ \JX=0 \vee \eta=0 $ \textsuperscript{(6)}
  \end{minipage} 
  \end{tabular} 

&
\begin{tabular}{l}
\begin{minipage}[t]{0.2\textwidth}
    \begin{itemize}
    \item $\Se_{X} \nearrow$
    \item $\Sp_{X} \nearrow$
    \item$\Se_{Z_{1}}\searrow$
    \item $\Sp_{Z_{1}}\searrow$
    \item $\eta \nearrow$
    \end{itemize}
  \end{minipage} \end{tabular}    &    $\eta \xi+(1-\eta)\varepsilon \leq 0$       &   \begin{tabular}{c} \begin{minipage}[p]{0.3\textwidth} \centering
  
$\eta \xi+(1-\eta)\varepsilon$ \\ $>$ \\ $(1-\eta) (1-\Sp_{Z_{1}})\JX$   \end{minipage} \end{tabular}                 \\
                 &            &      &     &           &                    \\
$\mathrm{CRS\_A}$ & Always     &  
  \begin{tabular}{l} 
    \begin{minipage}[t]{0.2\textwidth} \centering
     $\Se_{Z_{1}}=1 \wedge \Se_{Z_{2}}=1$\\ $\vee$ \\ $ \JX=0 \vee \eta=0 $ \textsuperscript{(4)}
  \end{minipage} 
  \end{tabular} 

    &  \begin{tabular}{l} \begin{minipage}[t]{0.2\textwidth}
    \begin{itemize}
    \item $\Se_{X} \nearrow$
    \item $ \Sp_{X} \nearrow$
    \item$\Se_{Z_{1}}\searrow$
    \item $\Sp_{Z_{1}}\searrow$ 
    \item$\Se_{Z_{2}}\searrow$
    \item $\Sp_{Z_{2}}\searrow$
    \item $\eta \nearrow$
    \end{itemize}
  \end{minipage} \end{tabular}   &     $\eta \Se_{Z_{2}} \xi+(1-\eta) (1-\Sp_{Z_{2}}) \varepsilon\leq0$        &    \begin{tabular}{c} \begin{minipage}[p]{0.3\textwidth} \centering $\eta \Se_{Z_{2}} \xi+(1-\eta) (1-\Sp_{Z_{2}}) \varepsilon$ \\ $>$ \\ $\eta  (1-\Se_{Z_{1}} \Se_{Z_{2}}) \JX$   \end{minipage} \end{tabular}          \\
                  &            &      &                &     &               \\
$\mathrm{CRS\_O}$ & Always     &   
  \begin{tabular}{l} 
    \begin{minipage}[t]{0.2\textwidth} \centering
     $\Se_{Z_{1}}=1 \vee \Se_{Z_{2}}=1$\\ $\vee$ \\ $ \JX=0 \vee \eta=0 $ \textsuperscript{(4)}
  \end{minipage} 
  \end{tabular} 
&  \begin{tabular}{l} \begin{minipage}[t]{0.2\textwidth}
    \begin{itemize}
    \item $\Se_{X} \nearrow $
    \item $\Sp_{X} \nearrow$
    \item$\Se_{Z_{1}}\searrow$
    \item $\Sp_{Z_{1}}\searrow$
    \item $\Se_{Z_{2}}\searrow$
    \item $\Sp_{Z_{2}}\searrow$
    \item $\eta \nearrow$
    \end{itemize}
  \end{minipage} \end{tabular}   &       $\eta (1-\Se_{Z_{2}}) \xi  +(1-\eta) \Sp_{Z_{2}} \varepsilon \leq 0$         &       \begin{tabular}{c} \begin{minipage}[p]{0.3\textwidth} \centering $\eta (1-\Se_{Z_{2}}) \xi  +(1-\eta) \Sp_{Z_{2}} \varepsilon$ \\ $ >$ \\ $\eta(1-\Se_{Z_{1}})(1-\Se_{Z_{2}})\JX$  \end{minipage} \end{tabular}         \\
                &            &      &     &                &               \\ \hline
\end{tabular}
}
}
\vspace{1ex}
\raggedright{\textsuperscript{(5)}} Any of the listed conditions contributes to decrease $\dSpM$ ($\dSpM \searrow$).\\
\raggedright{\textsuperscript{(6)}} $\JX=0$ and $\eta=0$ are unreasonable conditions, since $\JX=0$ implies a valueless index test, and $\eta=0$, i.e., a prevalence of zero, means that no subject in the population has the target condition. In fact, we assume $\JX>0$ and $\eta \in ]0,1[$.  
\end{table}

All the performance methods covered in our work lead to biased estimates of the accuracy measures under various circumstances. Besides the insight we can gain towards a specific case, based on the \See and \Spp theoretical expressions, we can also derive certain generic conclusions. We summarise some of these findings, regarding IGS, CRS\_A, and CRS\_O, in Tables~\ref{table:SeDeviations} and \ref{table:SpDeviations}. The findings derived from the analytical expressions of DA and LCM are not summarised in the same way due to the increased complexity of the expressions.


\section{Latent class model}
\label{sec:LCM}

In the absence of a gold standard, latent class models are widely used to estimate diagnostic tests performance measures, such as sensitivity and specificity, as well as the prevalence. In this context, a widely used LCM admits a binary latent variable, $Y$, such as defined in Subsection \ref{subsec2.1:notation}, whose categories are called latent classes and indicate the status regarding the condition of interest. Furthermore, the LCM admits manifest variables $X_{i}$ ($i=1,...,p$), that express the outcomes of $p$ diagnostic tests, assuming the value $1 \textrm{ if the $i$-th diagnostic test is positive}$ and $ 0 \textrm{ otherwise}$. The simplest LCM assumes the HCI previously mentioned, which means that the test results are independent conditional on the status of the condition of interest.
\\
Given $\Se_{i}=\Se_{X_{i}}=P(X_{i}=1|Y=1)$, $\Sp_{i}=\Sp_{X_{i}}=P(X_{i}=0|Y=0)$, with $i=1,2,3$, and $\eta=P(Y=1)$, then, for $p=3$, we have 
\begin{equation} \label{eq:LCM}
P(X_{1}=x_{1},X_{2}=x_{2},X_{3}=x_{3})=\eta \prod_{j=1}^{3} \Se_{j}^{x_{j}}(1-\Se_{j})^{1-x_{j}} + (1-\eta) \prod_{j=1}^{3} \Sp_{j}^{1-x_{j}}(1-\Sp_{j})^{x_{j}}.
\end{equation}
\\
The setup defined in Subsection \ref{subsec2.2:setup}, and used throughout the paper, is transposable to the LCM context, if we admit that the tests $X_{i}$ ($i=1,...,3$) correspond, respectively, to the test under study, $X$, and to the imperfect diagnostic tests, $Z_{1}$ and $Z_{2}$. 
\\
Next we will present analytical expressions for $\Se_{i}$, $\Sp_{i}$, and $\eta$ corresponding to the LCM under the HCI, as discussed in the literature \cite{Pepe2007,Formann08}. 
\\
Let us define
\begin{align}
& p_{123}= P^{\mathrm{HCI}}(X_{1}=1,X_{2}=1,X_{3}=1)=\eta \Se_{1} \Se_{2} \Se_{3}+(1-\eta) (1-\Sp_{1}) (1-\Sp_{2}) (1-\Sp_{3}), \label{eq:p123_LCM} \\
& p_{i}=P^{\mathrm{HCI}}(X_{i}=1)=\eta \Se_{i} + (1-\eta) (1-\Sp_{i}),  \nonumber \\  
& p_{ij}=P^{\mathrm{HCI}}(X_{i}=1,X_{j}=1)=\eta \Se_{i} \Se_{j} + (1-\eta) (1-\Sp_{i}) (1-\Sp_{j}), \nonumber \\  
& a_{ij}=p_{ij}-p_{i}p_{j}, \;\mathrm{with}\; \; i \neq j, \; \mathrm{and} \; \; i,j=1,2,3, \nonumber \\
& V=\frac{p_{123}-p_{12}p_{3}-p_{13}p_{2}-p_{23}p_{1}+2p_{1} p_{2} p_{3}}{\sqrt{a_{12} a_{13} a_{23}}}. \nonumber 
\end{align}
\\
For the sensitivities, we have
\begin{align} \label{Se_est_LCM}
\Se_{i}^{\mathrm{LCM}}=p_{i}+\sqrt{\frac{a_{ij} a_{ik}}{a_{jk}}} \sqrt{\frac{1-\eta}{\eta}},
\end{align}  
\\
where $i \neq j, \; i \neq k, \; j \neq k,\; \mathrm{and} \; \; i,j,k=1,2,3$,
\\
\\
and for the specificities, 
\begin{align} \label{Sp_est_LCM}
\Sp_{i}^{\mathrm{LCM}}=1-p_{i}+\sqrt{\frac{a_{ij} a_{ik}}{a_{jk}}} \sqrt{\frac{\eta}{1-\eta}}.
\end{align}  
\\
Regarding the two preceding expressions, \eqref{Se_est_LCM} and \eqref{Sp_est_LCM}, in each case the expression presented is chosen between two solutions. For example, we have $\Se_{i}^{\mathrm{LCM}}=p_{i}\pm \sqrt{\frac{a_{ij} a_{ik}}{a_{jk}}} \sqrt{\frac{1-\eta}{\eta}}$. The choice follows the reasonable assumption that the true positive-rate is at least as large as the false positive rate, i.e., $\Se_{i} \geq (1-\Sp_{i})$ \cite{Pepe2007}.
\\
It follows that
\begin{align} \label{eta_est_LCM}
\eta^{\mathrm{LCM}}=\frac{1}{2} \pm \sqrt{\frac{1}{4}-\frac{1}{4+V^{2}}},
\end{align}
and between these two solutions, the most commonly used is  $\eta^{\mathrm{LCM}}=\frac{1}{2} - \sqrt{\frac{1}{4}-\frac{1}{4+V^{2}}}$, since $\eta<0.5$ in most practical situations.
\\
We derived similar expressions under the HCI violation as defined in \eqref{eq:SetupDep}. Let us recall that we denote $\xi=cov(X_{1},X_{2}|Y=1)=cov(X,Z_{1}|Y=1)$ and $\varepsilon=cov(X_{1},X_{2}|Y=0)=cov(X,Z_{1}|Y=0)$. We will add the subscript $\overline{\mathrm{HCI}}$ to indicate that a certain expression concerns the HCI violation case.
\\
In this context,
\begin{align}
& P^{\overline{\mathrm{HCI}}}(X_{1}=1,X_{2}=1|Y=1)=\Se_{1} \Se_{2}+\xi, \nonumber \\
& P^{\overline{\mathrm{HCI}}}(X_{1}=1,X_{2}=1|Y=0)=(1-\Sp_{1}) (1-\Sp_{2})+ \varepsilon, \nonumber \\
& P^{\overline{\mathrm{HCI}}}(X_{1}=1,X_{2}=1,X_{3}=1|Y=1)=\Se_{1} \Se_{2} \Se_{3}+\Se_{3} \xi , \nonumber \\
& P^{\overline{\mathrm{HCI}}}(X_{1}=1,X_{2}=1,X_{3}=1|Y=0)=(1-\Sp_{1}) (1-\Sp_{2}) (1-\Sp_{3})+ (1-\Sp_{3}) \varepsilon. \nonumber
\end{align} 
\\
Based on these reformulated expressions and given the formulas in (\ref{eq:p123_LCM}), we obtain:
\begin{align}
& p_{123}^{\overline{\mathrm{HCI}}}=p_{123}+ \eta \Se_{3} \xi + (1-\eta) (1-\Sp_{3}) \varepsilon , \label{eq:p123_LCM2} \\
& p_{12}^{\overline{\mathrm{HCI}}}=p_{12} +  \eta \xi + (1-\eta)  \varepsilon, \nonumber \\
& p_{i3}^{\overline{\mathrm{HCI}}}=p_{i3}, \; \text{, because} \; X_{i} \bigCI X_{3} \mid Y=y, \;\text{with} \; y=0,1, \; \;\text{for} \; i=1,2,\nonumber \\
& p_{i}^{\overline{\mathrm{HCI}}}=p_{i}, \; \; i=1,2,3. \nonumber 
\end{align} 
\\
Under the HCI violations, the sensitivities\footnote{Here we present only one of the two possible solutions, based on the rationale previously explained.} emerge as 
\begin{align}
\Se_{1}^{\overline{\mathrm{LCM}}} & =p_{1} + \sqrt{\Big(\frac{1-\eta}{\eta}\Big) \frac{a_{12} a_{13}}{a_{23}} - (1-\eta) \frac{a_{13}}{a_{23}} \xi- \frac{(1-\eta)^2}{\eta} \frac{a_{13}}{a_{23}} \varepsilon }, \label{Se1_est_LCM} \\
\Se_{2}^{\overline{\mathrm{LCM}}} & =p_{2} + \sqrt{\Big(\frac{1-\eta}{\eta}\Big) \frac{a_{23} a_{12}}{a_{13}} - (1-\eta) \frac{a_{23}}{a_{13}} \xi- \frac{(1-\eta)^2}{\eta} \frac{a_{23}}{a_{13}} \varepsilon },  \label{Se2_est_LCM} \\
\Se_{3}^{\overline{\mathrm{LCM}}} & =p_{3} + \sqrt{\Big(\frac{1-\eta}{\eta}\Big) \frac{a_{13} a_{23}}{a_{12}-\eta \xi -(1-\eta) \varepsilon}}.  \label{Se3_est_LCM}
\end{align}
From the expression for $p_{i}$ ($i=1,2,3$), among Equations (\ref{eq:p123_LCM}), which is valid both under the HCI and under its violation, we can establish the following relation between $\Sp_{i}$ and $\Se_{i}$
\begin{align}
\Sp_{i}=1-\frac{1}{1-\eta}(p_{i}-\eta \Se_{i}).  \label{relSe_Sp}
\end{align}

It follows that the specificities according to the LCM under HCI violation are given by
\begin{align}
\Sp_{1}^{\overline{\mathrm{LCM}}} & =1 - p_{1} + \sqrt{\Big(\frac{\eta}{1-\eta}\Big) \frac{a_{12} a_{13}}{a_{23}} - \frac{\eta^2}{1-\eta} \frac{a_{13}}{a_{23}} \xi -\eta \frac{a_{13}}{a_{23}} \varepsilon }, \label{Sp1_est_LCM} \\
\Sp_{2}^{\overline{\mathrm{LCM}}} & =1 - p_{2} + \sqrt{\Big(\frac{\eta}{1-\eta}\Big) \frac{a_{12} a_{23}}{a_{13}} - \frac{\eta^2}{1-\eta} \frac{a_{23}}{a_{13}} \xi - \eta \frac{a_{23}}{a_{13}} \varepsilon },  \label{Sp2_est_LCM} \\
\Sp_{3}^{\overline{\mathrm{LCM}}} & =1 - p_{3} + \sqrt{\Big(\frac{\eta}{1-\eta}\Big) \frac{a_{13} a_{23}}{a_{12}-\eta \xi -(1-\eta) \varepsilon}}.  \label{Sp3_est_LCM}
\end{align}
\\
Introducing the expressions \eqref{Se_est_LCM} and \eqref{Sp_est_LCM}, obtained for $\Se_{i}^{\mathrm{LCM}}$  and $\Sp_{i}^{\mathrm{LCM}}$, with $i=1,2,3$, into the expression \eqref{eq:p123_LCM2}, derived for $p_{123}^{\overline{\mathrm{LCM}}}$ leads to
\begin{align}
& p_{123}^{\overline{\mathrm{LCM}}}=p_{1}p_{2}p_{3}\Big(1+a_{12}+a_{23}+a_{13}+\sqrt{a_{12} a_{13} a_{23}} \Big\{\sqrt{\frac{1-\eta}{\eta}}-\sqrt{\frac{\eta}{1-\eta}} \Big\} \Big)+ \eta \Se_{3} \xi + (1-\eta) (1-\Sp_{3}) \varepsilon. \label{eq:p123_LCM2b}
\end{align}
\\
If we define
\begin{align}
V_{1}&=p_{123}^{\overline{\mathrm{LCM}}}-p_{1}p_{2}p_{3}(1+a_{12}+a_{23}+a_{13}), \\
V_{2}&=p_{1}p_{2}p_{3} \sqrt{a_{12} a_{13} a_{23}}.
\end{align}
\\
then we obtain the expression
\begin{align}
V_{2} \Big\{\sqrt{\frac{1-\eta}{\eta}}-\sqrt{\frac{\eta}{1-\eta}} \Big\} +\eta p_{3} (\xi - \varepsilon) + \sqrt{\eta (1-\eta)\frac{a_{13} a_{23}}{a_{12}-\eta \xi -(1-\eta) \varepsilon}} (\xi - \varepsilon)= V_{1} - p_{3} \varepsilon.    \label{eq:p123_LCM2c}
\end{align}
\\
If we admit $\xi=\varepsilon$, i.e., $cov(X,Z_{1}|Y=1)=cov(X,Z_{1}|Y=0)$, which means that the conditional covariances between $X$ and $Z_{1}$ are  the same, then the previous expression simplifies into
\begin{align}
V_{2} \Big\{\sqrt{\frac{1-\eta}{\eta}}-\sqrt{\frac{\eta}{1-\eta}} \Big\}= V_{1} - p_{3} \varepsilon,    \label{eq:p123_LCM2d}
\end{align}
\\
leading to
\begin{align}
\eta=\frac{1}{2} \pm \sqrt{\frac{1}{4}-\frac{1}{4+W^{2}}}, \textrm{ where } W=\frac{V_{1}-p_{3} \varepsilon}{V_{2}}.    \label{p123_LCM2d}
\end{align}
\\
As explained earlier in the text, since typically $\eta<0.5$ , we opt for the solution 
\begin{align}
\eta^{\overline{\mathrm{LCM}}}=\frac{1}{2} - \sqrt{\frac{1}{4}-\frac{1}{4+W^{2}}}.    \label{eq:eta_nHCI}
\end{align}

Latent class models yield unbiased estimates of the parameters when the underlying assumptions hold, which means that the LCM \See and \Spp expressions derived earlier lead to the true population parameters and no deviations can be quantified, when the population and the LCM coincide regarding the dependence structure. In order to evaluate potential deviations arising from the use of LCM, we model situations in which an incorrect LCM for a specific population is applied. In fact, we admit two scenarios:
\begin{enumerate}
    \item LCM under HCI is used, although this assumption is invalid in the population, which verifies local dependence according to \eqref{eq:SetupDep}. This scenario, which we designate LCM(HCI), summarises situations where the population features conditional dependencies between the index test and one of the imperfect references, which are ignored by the model adjusted to the data, since the LCM assumes the HCI. Hence, in this case we quantify the effect of using the LCM incorrectly assuming the HCI. 
    \item LCM assuming the violation of HCI, as defined in \eqref{eq:SetupDep}, is adopted, although conditional independence between the tests is valid in the population. This scenario, denominated $\mathrm{LCM(\overline{\mathrm{HCI}})}$, addresses a situation where an unnecessary complex model is adjusted to the population. Indeed, the HCI is valid in the population, but the model introduces two unnecessary parameters, $\xi=cov(X_{1},X_{2}|Y=1)$ and $\varepsilon=cov(X_{1},X_{2}|Y=0)$, to model nonexistent local dependencies.
\end{enumerate}

The complexity of the \See and \Spp expressions makes it difficult to  understand how these vary with the prevalence, diagnostic accuracy measures, and dependencies between the tests. In order overcome this problem and identify \See and \Spp patterns of variation, deviations for specific values with practical relevance can be calculated and visualised using an \R Shiny application made available online to the reader. 
\\


\section{Admissible values for the conditional covariances}
\label{sec:Covariances_limits}

The parameters $\xi$ and $\varepsilon$ model local dependencies between the diagnostic test $X$ and one of the imperfect reference tests, $Z_{1}$. In this Section, we report the ranges of acceptable values for $\xi$ and $\varepsilon$ required by the different methods and cases covered in the paper.

For a start, we must ensure that the values of $\xi$ and $\varepsilon$ lead to $0 \leq P(X,Z_{1}|Y=y) \leq 1$, with $y=0,1$. Thus, regardless of the method M, assuming the violation of HCI in the population, we must impose
\begin{align}
& \max(-\SeX \SeZ,-(1-\SeX)(1-\SeZ)) \leq \xi \leq min(\SeX,\SeZ)-\SeX \SeZ, \text{ and} \label{eq:restr1} \\
& \max(-\SpX \SpZ,-(1-\SpX)(1-\SpZ)) \leq \varepsilon \leq min(\SpX,\SpZ)-\SpX \SpZ. \nonumber
\end{align}

For the particular case of the LCM, the analytical expressions for $\See$, $\Spp$, and $\eta$ may lead to values that lie outside the admissible interval $[0,1]$, a problem previously reported in the literature \citep[e.g.,][]{Formann08}. To workaround this difficulty, instead of restricting the values of $\xi$ and $\varepsilon$ to ranges for which the theoretical expressions lie within $[0,1]$, we impose  $\Se_{i}^{\mathrm{LCM}}=\min(\Se_{i}^{\mathrm{LCM}},1) \wedge \Se_{i}^{\mathrm{LCM}}=\max(\Se_{i}^{\mathrm{LCM}},0)$, with $i=1,2,3$, and force the same for $\Spp$ and $\eta$, for the LCM both under the HCI and the HCI violation. 

When the population and the LCM do not match regarding the HCI, corresponding to the scenarios $\mathrm{LCM(HCI)}$ and $\mathrm{LCM(\overline{\mathrm{HCI}})}$, defined in Section \ref{sec:LCM}, additional restrictions on $\xi$ and $\varepsilon$ are needed. In fact, the analytical expressions of $\See$, $\Spp$, and $\eta$ according to the LCM include square roots, whose radicands must be non-negative. Specifically for $\mathrm{LCM(HCI)}$ and $\mathrm{LCM(\overline{\mathrm{HCI}})}$, in order to have non-negative radicands in the analytical formulas, the values of $\xi$ and $\varepsilon$ must verify the conditions:
\begin{align}
& \eta \xi + (1-\eta) \varepsilon \geq -\eta (1-\eta) (1-\Se_{1}-\Sp_{1})(1-\Se_{2}-\Sp_{2}), \text{ for } \mathrm{LCM(HCI)}, \text{ and} \label{eq:restr2} \\
& \eta \xi + (1-\eta) \varepsilon \leq \eta (1-\eta) (1-\Se_{1}-\Sp_{1})(1-\Se_{2}-\Sp_{2}), \text{ for } \mathrm{LCM(\overline{\mathrm{HCI}})}. \nonumber
\end{align}

In the \R Shiny application, for every set of parameters specified by the user, the valid limits of $\xi$ and $\varepsilon$ are calculated, and the plots reflect these changes. In some situations, the ranges of possible values for the different methods do not coincide, whereby the various methods represented in the same plot may correspond to different ranges of values in the horizontal axis.




\section{\See and \Spp deviations motivated by a practical problem} 
\label{sec:application}

We now illustrate the practical utility of the theoretical results, by examining the deviations determined for a particular setting, defined by a list of sensitivities and specificities for the three diagnostic tests ($X$, $Z_{1}$, and $Z_{2}$), prevalence of the condition, $\eta$, and conditional covariances, $\xi$ and $\varepsilon$. Besides these fixed values, realistic variation intervals, that mimic the researcher's uncertainty about the parameters, are also proposed. It is then possible to investigate the effect of factors we deem relevant (accuracy of the tests, prevalence, and  conditional dependence between the tests) on $\Delta {\rm Se}^{\mathrm{M}}$ and $\Delta {\rm Sp}^{\mathrm{M}}$, by varying the factors within ranges of plausible values. 

The fixed values and variation intervals were chosen according to studies concerning \emph{C. Trachtomatis} or infectious diseases such as tuberculosis or pneumonia (see \cite{Schiller2015} and references therein). Hence, we admit the following fixed populational values:  
\begin{itemize}
\item Equal $\Se$ and $\Sp$ for the index test, $\SeX=\SpX=0.90$;
\item Equal $\See$ for both imperfect reference tests, $\SeZ=\SeZZ=0.60$;
\item Equal $\Sp$ for both imperfect reference tests, $\SpZ=\SpZZ=0.95$;
\item Prevalence, $\eta=0.10$.
\end{itemize}

As for the variation intervals, we presume the following intervals of plausible values:  
\begin{itemize}
\item $\SeZ=\SeZZ$ take values in $(0.30,0.90)$, a wide range of values, spanning from low to high sensitivities;
\item $\SpZ=\SpZZ$ vary within $(0.90,1)$, i.e., we admit high values for the specificities;
\item Prevalence $\eta$ assumes values in $(0.05,0.30)$, corresponding to low prevalences;
\item Conditional covariances range from $\xi=\varepsilon=0$, corresponding to the HCI, to strong conditional dependence, with $\xi$ and $\varepsilon$ assuming both positive and negative values within the range of values acceptable according to \eqref{eq:restr1} and \eqref{eq:restr2}.
\end{itemize}

In Subsection~\ref{subsec:applicationHCI} we study the \See and \Spp deviations in the case of conditionally independent tests. The methods IGS, $\mathrm{CRS\_A}$, $\mathrm{CRS\_O}$, and DA are addressed. The deviations in the context of conditionally dependent tests are investigated in Subsection~\ref{subsec:applicationHCIviolation}, in which case the LCM assuming the HCI is also addressed. Finally, in Subsection~\ref{subsec:applicationLCM}, we explore the deviations resulting from the use of LCM assuming inappropriate dependence structures, i.e., $\mathrm{LCM(HCI)}$ and $\mathrm{LCM(\overline{HCI})}$.

\subsection{Deviations under the HCI} 
\label{subsec:applicationHCI}

\begin{figure}
\centering
\includegraphics[width=0.9\textwidth]{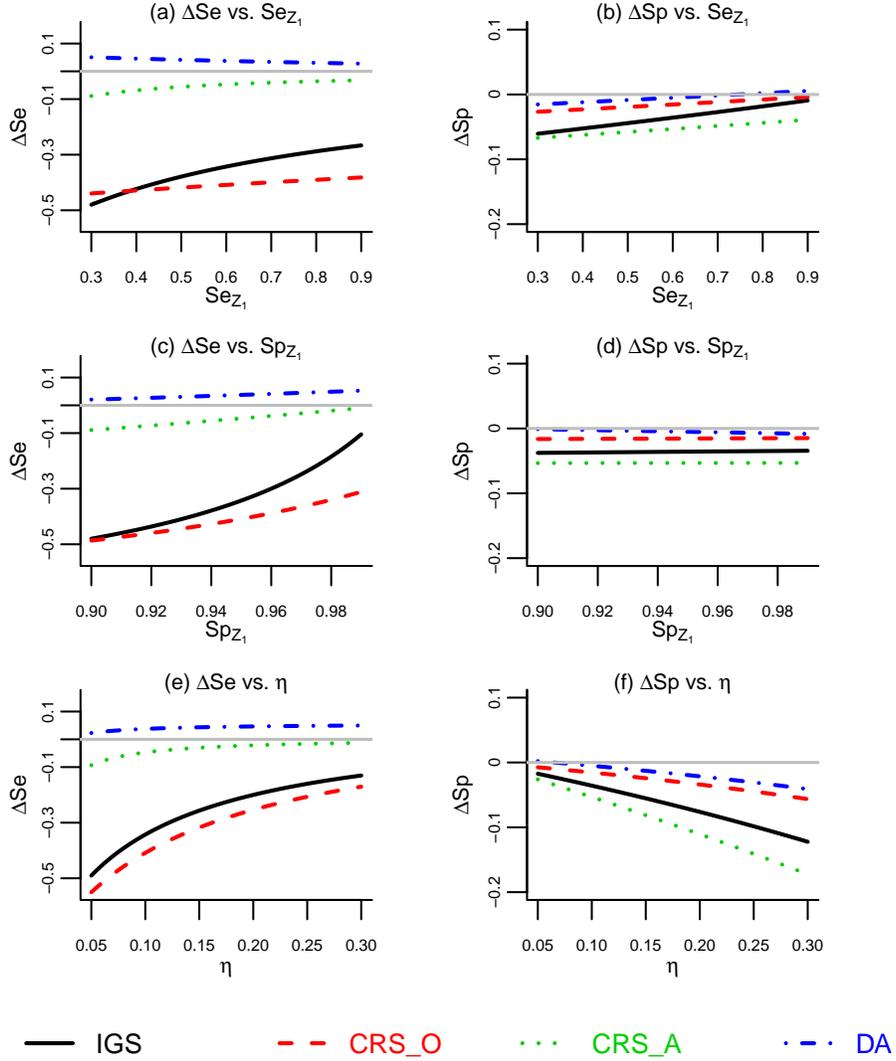}
\caption{Plots of $\Delta {\rm Se}^{\mathrm{M}}$ (left panel, (a), (c), (e)) and $\Delta {\rm Sp}^{\mathrm{M}}$ (right panel, (b), (d), (f)) versus $\Se_{Z_{1}}$ (upper panel, (a) and (b)), $\Sp_{Z_{1}}$ (middle panel, (c) and (d)), and the prevalence $\eta$ (lower panel, (e) and (f)), assuming the HCI. The methods $\mathrm{M}$ under comparison are IGS (solid black line), $\mathrm{CRS\_O}$ (dashed red line), $\mathrm{CRS\_A}$ (dotted green line), and $\mathrm{DA}$ (dashed-dotted blue line).}\label{f1}
\end{figure}

Figure~\ref{f1} presents the plots of the deviations $\Delta {\rm Se}^{\mathrm{M}}$ and $\Delta {\rm Sp}^{\mathrm{M}}$, for the $\mathrm{M}$ methods IGS, $\mathrm{CRS\_O}$, $\mathrm{CRS\_A}$, and $\mathrm{DA}$, varying with the $\See$ and $\Spp$ of the imperfect reference test $Z_{1}$, and also with the prevalence, assuming local independence between the tests ($\xi=\varepsilon=0$). These plots show that:
\begin{itemize}
\item $\Spp$ deviations, $\Delta {\rm Sp}^{\mathrm{M}}$, tend to be slimmer than $\See$ deviations, $\Delta {\rm Se}^{\mathrm{M}}$, (Figure~\ref{f1}, right plots against left, noting the differences in y-axis scales), implying smaller bias in $\Spp$ estimation than $\Se$; 
\item $\Delta {\rm Sp}$ vary slightly more with the prevalence, $\eta$, (Figure~\ref{f1} (f)), than with the accuracy of the reference $Z_{1}$, $\Se_{Z_{1}}$ and $\Sp_{Z_{1}}$ (Figure~\ref{f1} (b) and (d));
\item Figure~\ref{f1} (a), (c), and (e) show that $\See$ deviations are always negative for the IGS, $\mathrm{CRS\_O}$, and $\mathrm{CRS\_A}$. The $\See$ according to these three methods approach the true $\See$, as $\Se_{Z_{1}}$ or  $\Sp_{Z_{1}}$ increases;
\item On the contrary, $\Delta {\rm Se}^{\mathrm{DA}}$ assume positive values and slightly decrease with $\Se_{Z_{1}}$ (Figure~\ref{f1} (a)), but thinly increase with $\Sp_{Z_{1}}$ and $\eta$ (Figure~\ref{f1} (c) and (e));
\item As expected, the absolute value of the $\See$ deviations, except for $\Delta {\rm Se}^{\mathrm{DA}}$, decrease with the prevalence (Figure~\ref{f1} (e)) and the absolute value of the $\Spp$ deviations increase (Figure~\ref{f1} (f)).
\item $\Delta {\rm Se}^{\mathrm{DA}}$ range from values near $0.02$ to $0.05$, while $\Delta {\rm Se}^{\mathrm{CRS\_A}}$ span from approximately $-0.09$ to $-0.01$. Thus, for this particular setting, $\mathrm{CRS\_A}$ and $\mathrm{DA}$ seem preferable methods to estimate the $\See$ under HCI;
\item Regarding the $\Spp$, however, $\mathrm{CRS\_A}$ leads to estimates farther away from the true $\Spp$, while $\mathrm{DA}$ and $\mathrm{CRS\_O}$ based $\Spp$ are nearer the target (Figure~\ref{f1} (b), (d), and (f)).
\end{itemize}

Based on the theoretical deviations presented in Figure~\ref{f1}, we aim to determine which method is preferable. We start with $\dSeM$ under the HCI (Figure~\ref{f1} (a), (c), (e)): 
\begin{itemize}
\item DA overestimates $\SeX$, while the remaining methods underestimate. 
\item Both $\mathrm{CRS\_O}$ and IGS strongly underestimate $\SeX$. $\mathrm{CRS\_O}$ is the worst of the two, except for $\SeZ \in [0.3,0.4]$, in which case the IGS leads to estimates farther away from $\SeX$. 
\item By contrast, $\mathrm{CRS\_A}$ and DA are the methods that produce estimates closer to the truth, although $\mathrm{CRS\_A}$ underestimates the $\SeX$, while DA overestimates.
\end{itemize}
Furthermore, we can compare $\mathrm{CRS\_A}$ and DA, which are the best methods for \See estimation in this case, based on the effect of the different factors we explored:
\begin{itemize}
\item $\dSeM $ vs. $\SeZ$ (Figure~\ref{f1} (a)): As $\SeZ$ increases, both $|\dSeCRSand|$ and $|\dSeDA|$ decrease. For all values of $\SeZ$, $|\dSeDA|<|\dSeCRSand|$, i.e., DA leads to estimates closer to the true $\SeX$. DA is noticeable better than $\mathrm{CRS\_A}$, for lower values of $\SeZ$, but for higher values of $\SeZ$, the two methods lead to deviations from $\SeX$ similar in absolute value, even if DA overestimates and $\mathrm{CRS\_A}$ underestimates $\SeX$.  
\item $\dSeM $ vs. $\SpZ$ (Figure~\ref{f1} (c)): The curves of $\dSeCRSand$ and $\dSeDA$ plotted against $\SpZ$ have both positive slope. Given that $\dSeDA>0$, it follows that the overestimation of $\SeX$ increases with $\SpZ$, i.e., the estimates get farther away from the true $\SeX$. By contrast, since $\dSeCRSand<0$, then $\mathrm{CRS\_A}$ leads to estimates of $\SeX$ closer to the true values as $\SpZ$ increases. Accordingly, DA is preferable to $\mathrm{CRS\_A}$ for lower values of $\SpZ$ and vice versa.
\item $\dSeM $ vs. $\eta$ (Figure~\ref{f1} (e)): The variation of $\dSeCRSand$ and $\dSeDA$ against $\eta$ is similar to the pattern described in the preceding item for the variation against $\SpZ$. Accordingly, we can replicate the previous comments, stating that DA delivers better estimates of $\SeX$ for lower prevalences and $\mathrm{CRS\_A}$ for higher.    
\end{itemize}
We now explore the case of $\dSpM$ under the HCI (Figure~\ref{f1} (b), (d), (f)):
\begin{itemize}
\item The relative ordering of the curves replicates the one obtained for the $\dSeM$. However, in this case, $\dSpM<0$ for all the methods, including DA, which assumes negative values except for $\SeZ > 0.75$ or $\eta < 0.06$. 
\item The two best methods for $\SpX$ estimation in this case are DA and $\mathrm{CRS\_O}$.
\end{itemize}
Regarding the impact of $\SeZ$, $\SpZ$, and $\eta$, we notice that DA leads to $\SpX$ estimates closer to the true value, with the following exceptions:
\begin{itemize}
\item $\mathrm{CRS\_O}$ is better than DA for $\SeZ > 0.88$, although the difference is of the order of magnitude of $10^{-3}$ (Figure~\ref{f1} (b)).
\item $\mathrm{CRS\_O}$ is slightly better for $\eta<0.03$ (Figure~\ref{f1} (f)). 
\end{itemize}
In conclusion, combining the findings reached for $\dSeM$ and $\dSpM$, DA is the method that leads to better results, except for some limited cases.

\subsection{Deviations under the violation of HCI} 
\label{subsec:applicationHCIviolation}

\begin{figure}
\centering
\includegraphics[width=0.8\textwidth]{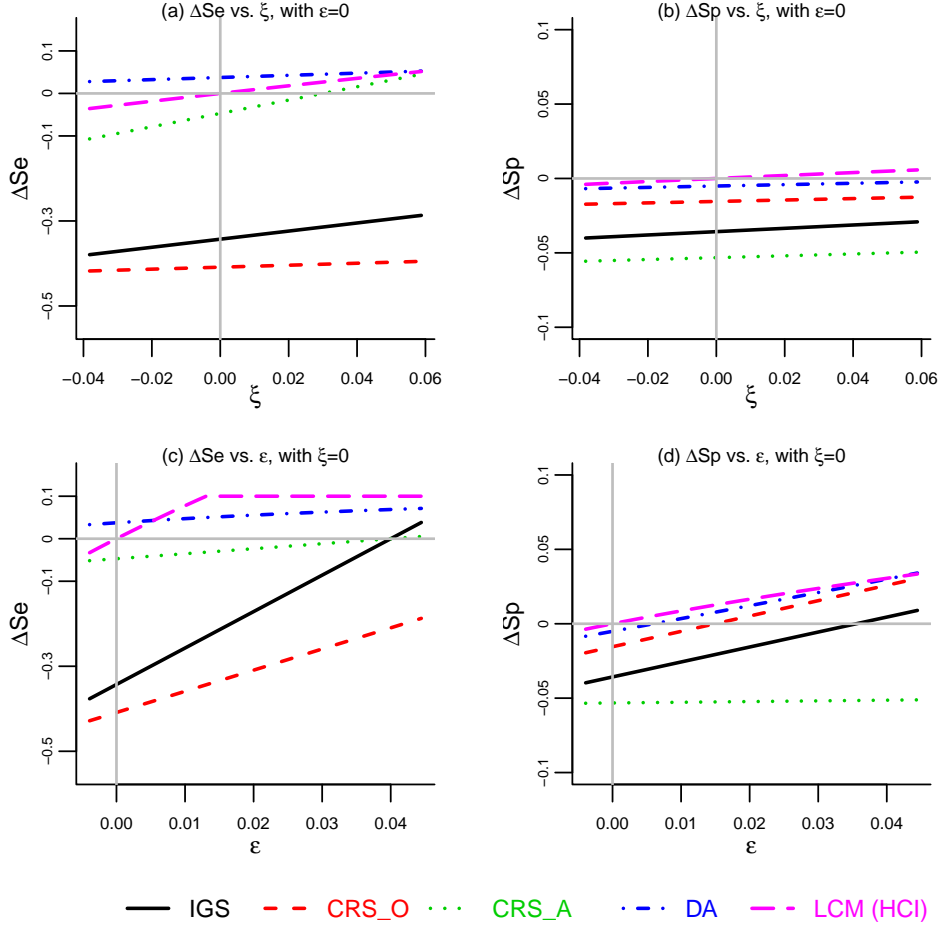}
\caption{Plots of $\dSenM$ (left panel, (a) and (c)) and $\dSpnM$ (right panel, (b) and (d)) versus the conditional covariances, $\xi=cov(X,Z_1|Y=1)$ (upper panel, (a) and (b)) and $\varepsilon==cov(X,Z_1|Y=0)$ (lower panel, (c) and (d)). The methods $\mathrm{M}$ under comparison are IGS (solid black line), $\mathrm{CRS\_O}$ (dashed red line), $\mathrm{CRS\_A}$ (dotted green line), $\mathrm{DA}$ (dashed-dotted blue line), and $\mathrm{LCM (HCI)}$ (long-dashed purple line).}\label{f2}
\end{figure}

Figure~\ref{f2} shows the plots of the deviations $\dSenM$ and $\dSpnM$ varying with the conditional covariances, $\xi$ and $\varepsilon$, for the methods $\mathrm{M}$, which are IGS, $\mathrm{CRS\_A}$, $\mathrm{CRS\_O}$, $\mathrm{DA}$, and the LCM assuming the HCI, when this assumption does not hold in the population, i.e., scenario $\mathrm{LCM(HCI)}$. Both positive and negative local dependencies are investigated, ergo $\xi$ and $\varepsilon$ may assume either positive or negative values. When the conditional covariance in one of the classes varies, conditional independence is assumed in the other class, which means that, when we plot $\dSenM$ and $\dSpnM$ against $\xi$, we admit $\varepsilon=0$, and vice versa. In the plots of Figure~\ref{f2} we notice that: 

\begin{itemize}
 \item The slopes of the lines are systematically positive in the four plots, except for the $\Delta {\rm Se}^{\mathrm{LCM(HCI)}}$, with positive slopes for lower values of $\varepsilon$, but zero slope for higher values (Figure~\ref{f2} (c)), as a result of truncation due to the constraint $\Se _{X}^{\mathrm{LCM(HCI)}} \leq 1$, resulting in $ \Delta {\rm Se}^{\mathrm{LCM(HCI)}} \leq 0.1$, since $\SeX=0.90$.
 \item Since the slopes of the deviations are positive, increasing $\xi$ or $\varepsilon$ leads to $\SenM$ and $\SpnM$ closer to the true values if the deviations are negative (e.g., IGS in Figure~\ref{f2} (a)), and $\SenM$ and $\SpnM$ farther away from the true values if the deviations are positive (e.g., DA in Figure~\ref{f2} (b)).
 \item $\dSpnM$ varies little with $\xi$ for all the methods (Figure~\ref{f2} (b)), and more with $\varepsilon$, except for $\mathrm{CRS\_A}$, almost unaffected by the changes of $\varepsilon$, as well as $\xi$ (Figure~\ref{f2} (d)).
  \item The slopes corresponding to $\dSpnM$ against $\xi$ are similar for all the methods (Figure~\ref{f2} (b)), and also for $\dSpnM$ against $\varepsilon$, with the exception of $\mathrm{CRS\_A}$ (Figure~\ref{f2} (d)).
  \item Steeper deviations are perceived for $\dSenM$ than for $\dSpnM$ (Figure~\ref{f2} (a) and (c) against (b) and (d), which have different y-axis scales).
  \item The changes in $\xi$ have a smaller effect in $\dSenM$ than the changes in $\varepsilon$, except for $\mathrm{CRS\_A}$ (Figure~\ref{f2} (a) against (c)).
 \item $\mathrm{CRS\_A}$,  IGS, and $\mathrm{LCM \, (HCI)}$ are the methods which show larger variations of $\dSenM$ against $\xi$ (Figure~\ref{f2} (a)). 
  \item The largest variations of $\dSenM$ with $\varepsilon$ (Figure~\ref{f2} (c)) are again obtained for IGS and $\mathrm{LCM \, (HCI)}$, but, in contrast with the previous case, for $\mathrm{CRS\_O}$.
\end{itemize}

As far as the violation of HCI goes, let us compare the methods in more detail, based on the variation of  $\dSenM$ and $\dSenM$ with $\xi$, under the assumption $\varepsilon=0$ (Figure~\ref{f2} (a), (b)):
\\
\begin{itemize}
\item Same as under the HCI, DA is the best method in some cases ($\xi < -0.032$). $\mathrm{CRS\_A}$ is preferable for $\xi > 0.021$, and for values of $\xi$ in between $\mathrm{LCM (HCI)}$ gives estimates of $\SeX$ closer to the true values.
\item $\mathrm{LCM (HCI)}$ estimates $\SpX$ slightly better than the remaining methods for all the valid values of $\xi$, other than $\xi>0.03$, in which case DA dimly surpasses $\mathrm{LCM (HCI)}$. 
\item Combining the results obtained for the estimation of $\SeX$ and $\SpX$, we can argue that $\mathrm{LCM (HCI)}$ is preferable for almost all values of $\xi$, despite the impact of $\xi$ on the $\mathrm{LCM (HCI)}$ estimates.
\end{itemize}

Comparing the methods regarding the effect on $\dSenM$ and $\dSpnM$ of $\varepsilon$, under the assumption $\xi=0$ (Figure~\ref{f2} (c), (d)):
\\
\begin{itemize}
\item The estimation of $\SeX$ is considerably affected by the changing $\varepsilon$, for all the methods. Among these, DA and $\mathrm{CRS\_A}$ are the ones that vary less.
\item $\mathrm{CRS\_A}$ is the only method that remains almost unchanged for different values of $\varepsilon$ regarding the estimation of $\SpX$. All the other methods are clearly affected, underestimating $\SpX$ for lower values of $\xi$ and overestimating for higher values.  
\item For different ranges of $\varepsilon$, in turn, the methods $\mathrm{LCM (HCI)}$, DA, $\mathrm{CRS\_O}$ or IGS lead to estimates of $\SpX$ closer to its true value. 
\item Taking into account both $\dSenM$ and $\dSpnM$ against $\varepsilon$, it follows that $\mathrm{LCM(HCI)}$ is the best method for lower values of $\varepsilon$, while IGS gives better results for higher values, and DA is preferable for a certain range in between.  
\end{itemize}

\subsection{Deviations for LCM assuming invalid dependence structures} 
\label{subsec:applicationLCM}

We explore the \See and \Spp deviations arising from the mismatch between the population and the LCM, in terms of conditional dependence between the tests. As detailed earlier in the end of Section \ref{sec:LCM}, we address the following two cases: (a) HCI invalid in the population, but a LCM under HCI is used to model it, i.e., an overly simplistic model for the population is adopted; (b) HCI valid in the population, but a LCM assuming conditional dependence is used, i.e., an unnecessarily complex model is adopted. The first case is designated ${\mathrm{LCM(HCI)}}$, and the second $\mathrm{LCM(\overline{\mathrm{HCI}})}$.

\begin{figure}
\centering
\includegraphics[width=0.95\textwidth]{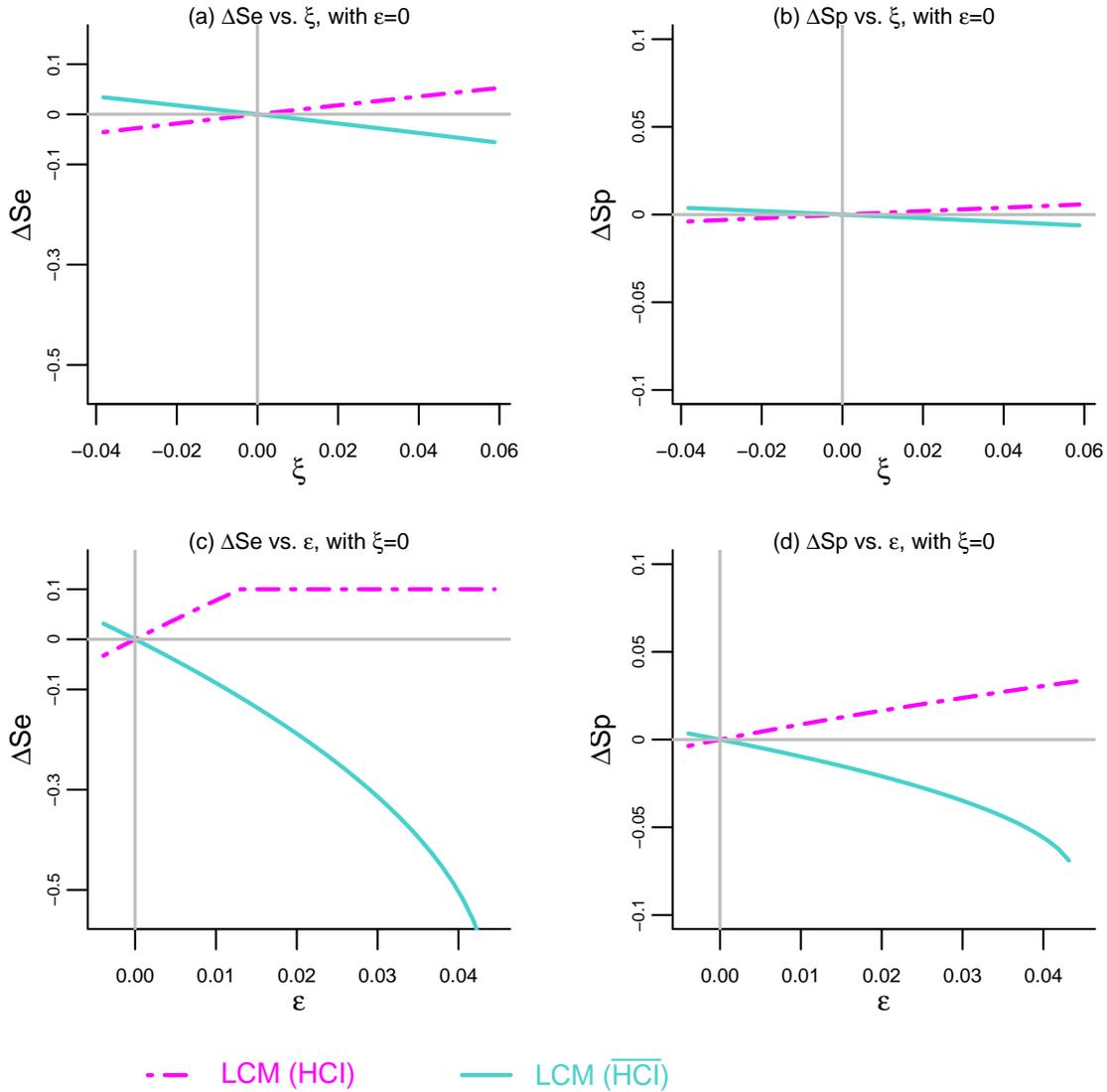}
\caption{Plots of $\dSenM$ (left panel, (a) and (c)) and $\dSpnM$ (right panel, (b) and (d)) versus the conditional covariances $\xi$ (upper panel, (a) and (b)) and $\varepsilon$ (lower panel, (c) and (d)). $M \in \{\mathrm{LCM (HCI)},\mathrm{LCM (\overline{HCI})} \}$, such that $\mathrm{LCM (HCI)}$ corresponds to applying a LCM under HCI to a population with conditionally dependent tests, whereas $\mathrm{LCM (\overline{HCI})}$ stands for the case in which a LCM assuming conditional dependence is adopted, but the HCI is valid in the population.}\label{f3}
\end{figure}

Figure~\ref{f3} exhibits the plots of the deviations $\dSenM$ and $\dSpnM$ against the covariances $\xi$ and $\varepsilon$, for $\mathrm{M} \in \{{\mathrm{LCM (HCI)}},\mathrm{LCM (\overline{\mathrm{HCI}})} \}$. The plots in Figure~\ref{f3} shows situations where conditional dependence in presumed in only one of the classes at a time. In fact, when the variation of $\xi$ is investigated, it is assumed that $\varepsilon = 0$ (upper panel, (a) and (b)), while under the variation of $\varepsilon$, we have $\xi=0$ (lower panel, (c) and (d).   

According to the plots in Figure~\ref{f3}:
\begin{itemize}
\item The changes in $\varepsilon$ have a greater effect in $\Delta {\rm Se}^{\mathrm{M}}$ and $\Delta {\rm Sp}^{\mathrm{M}}$ than the changes in $\xi$ (Figure~\ref{f3} (c) and (d) against (a) and (b)), i.e., the effect is higher when the conditional dependence is introduced in the class with highest frequency.
\item The impact of varying the conditional covariance, whether $\xi$ or $\varepsilon$, is greater in $\dSenM$ than $\dSpnM$ (Figure~\ref{f3} (a) and (c) against (b) and (d)).
\item $\mathrm{LCM (HCI)}$ and  $\mathrm{LCM (\overline{HCI})}$ lead to deviations in opposite directions.
\item For $\xi<0$, as well as for $\varepsilon<0$, $\mathrm{LCM (HCI)}$ leads to negative \See and \Spp deviations, whereas $\mathrm{LCM(\overline{HCI})}$ yields positive deviations. Inversely, for $\xi>0$ or $\varepsilon>0$, $\mathrm{LCM (HCI)}$ leads to positive \See and \Spp deviations, while $\mathrm{LCM(\overline{HCI})}$ results in negative deviations.
\item The deviations are null if $\xi=0$ in Figure~\ref{f3}, (a) and (b), or $\varepsilon=0$ in Figure~\ref{f3}, (c) and (d), corresponding to the case  $\xi=\varepsilon=0$ for both the population and the LCM, i.e., coherence between the population and the LCM regarding the conditional dependence.
\item $\Delta {\rm Se}^{\mathrm{LCM (HCI)}}$ and $\Delta {\rm Se}^\mathrm{LCM (\overline{HCI})}$ vs. $\xi$ have similar absolute values in Figure~\ref{f3} (a), just as $\Delta {\rm Sp}^{\mathrm{LCM (HCI)}}$ and $\Delta {\rm Sp}^\mathrm{LCM (\overline{HCI})}$ vs. $\xi$ have very close absolute values in Figure~\ref{f3} (b). 
\item The \See and \Spp deviations in Figure~\ref{f3} (c) and (d) display slightly larger absolute values for $\mathrm{LCM (\overline{HCI})}$ than for $\mathrm{LCM (HCI)}$, i.e., $\mathrm{LCM (\overline{HCI})}$ leads to \See and \Spp deviations farther away from the true values than $\mathrm{LCM (HCI)}$, for higher values of $\varepsilon$. In this case, the method that introduces inappropriate conditional dependence when the HCI is valid in the population, i.e., a needlessly complex model, leads to estimates farther away from the truth than the simplified model $\mathrm{LCM (HCI)}$. 
\end{itemize}


\section{Discussion}
\label{sec:discussion}

We propose a theoretical approach to investigate and compare some of the most undemanding and commonly used methods to evaluate the accuracy of a diagnostic test in the absence of a gold standard. We rely on analytical expressions for the deviations between the sensitivity and specificity according to each method, and the corresponding true values. The effect on these theoretical deviations of factors such as the tests accuracy, prevalence, and conditional dependence between the tests may be explored and visualised through the use of an interactive \R Shiny application, made available online along with this article.

We believe that our approach is of great practical utility, in spite of its theoretical nature. In fact, the researcher with a mere practical interest, may bypass the deduction of the mathematical expressions, and simply explore the formulas to gain further insight into the methods. The \R Shiny application allows the user to perform a sensitivity analysis, studying the magnitude and direction of the deviations relevant for a particular case of interest, based on ranges of values anticipated for the parameters. An additional merit of this approach is the removal of the confounding effects arising from the sampling scheme.

The methods addressed in this work may be organised into two groups. On the one hand, $\mathrm{IGS}$, $\mathrm{CRS\_A}$, $\mathrm{CRS\_O}$, and $\mathrm{DA}$ form a set of methods that evaluate the index test's accuracy based on the direct comparison with a single or multiple imperfect diagnostic tests. These methods can be simplified into a unified formulation as an IGS. On the other hand, the LCM stands for a different approach, in which the model combines results from multiple diagnostic tests to estimate the performance measures and the prevalence of the condition.

DA stands out among the methods in the first group. Underlying DA is the intuitive idea whereby a subset of the cases for which disagreement between the index test and the imperfect reference occurs, should be subject to an additional reference test to validate the outcome. It follows that the index test itself contributes to define the reference against which it is evaluated, a controversial strategy, harshly criticised in the literature. DA estimates are obtained without correct model specification, since the singularities of the DA procedure are omitted from the model. Imputation could prove promising to cope with this drawback and improve DA estimates. 

Regarding DA, we can also point out that it tends to overestimate the accuracy measures more frequently than the remaining methods in the group. In the biomedical context, the ethical implications associated with overestimating the accuracy measures are more serious than in the case of underestimation. However, if the direction of the bias could be overlooked, and only the magnitude mattered, then DA could be preferable in some cases, such as the one described in Subsection~\ref{subsec:applicationHCI}, when it provides closer estimates to the true values than the other methods. These findings may be transposed to fields where DA is also applied, but the overestimation of the performance is less alarming. 

CRS and LCM emerge in the literature as two of the most promising methods to evaluate diagnostic tests performance in the absence of a gold standard. An advantage of the CRS is that it offers the possibility of incorporating previous available information on the tests. In fact, the CRS estimates improve when the rule chosen to combine the tests follows from prior knowledge on their characteristics. The basic LCM does not integrate this kind of preliminary information, but it is also feasible to do so in the context of LCM, by introducing restrictions into the model or otherwise by adopting a Bayesian approach, whereby additional information can be incorporated through prior informative distributions. 

In our setup, we defined $\eta$ as the prevalence of the condition of interest. However, in practice, the experimental design may lead to samples that do not reflect the parent population regarding the prevalence. Since \See and \Spp deviations may vary considerably with $\eta$, the estimates obtained for these accuracy measures may change with the sampling schemes and the corresponding proportions of individuals with the condition in the sample. 

As future work, we intend to generalise our approach to cases that are not covered in the present work. Here we admit three diagnostic tests, but it would be pertinent to include more tests in the setup. Also, other evaluation methods should be explored under a similar approach, namely latent class models with random effects to accommodate conditional dependence between the tests, bayesian latent class models, and imputation approaches to LCM. Moreover, if the complexity of some of these models undermines the derivation of theoretical expressions for the \See and \Spp deviations, the problem may be approached via a simulation study.

\section{Acknowledgements}

This work has been supported by Funda\c{c}\~{a}o para a Ci\^{e}ncia e Tecnologia (FCT), Portugal, through the UID/Multi/04621/2013 and PTDC/EEI-TEL/5708/2014 projects. Ana Subtil is funded by the FCT PhD grant SFRH/BD/69793/2010.

\bibliographystyle{wileyj}  
\bibliography{refs}
\end{document}